\documentclass[structabstract]{aa}  
\usepackage{graphicx}
\usepackage{txfonts}
\usepackage{natbib}
\usepackage{booktabs}
\usepackage{upgreek}
\usepackage{setspace}
\usepackage{comment}
\usepackage{multirow}
\usepackage{float}
\usepackage{textcomp}
\usepackage{dcolumn}
\def \h2{\ifmmode{{\rm H}_2}\else{${\rm H}_2$}}

\def \vhel{\ifmmode{~V_{{\rm HEL}}}\else{~$V_{{\rm HEL}}$}\fi}
\def \mycn18 {MyCn~18}

\begin{document}
   \title{A morpho-kinematic and spectroscopic study of the bipolar nebulae: M~2--9, Mz~3, and Hen~2--104}

   \subtitle{}

   \author{N. Clyne
          \inst{1,2},
          S. Akras\inst{2}, W. Steffen\inst{3}, M. P. Redman\inst{1}, D. R. Gon\c{c}alves\inst{2}, E. Harvey\inst{1}}

   \institute{Centre for Astronomy, School of Physics, National University of Ireland Galway, University Road, Galway, Ireland \\
              \email{n.clyne1@nuigalway.ie, matt.redman@nuigalway.ie}
	\and
	{Observat\'orio do Valongo, Universidade Federal do Rio de Janeiro, Ladeira Pedro Antonio 43 20080-090 Rio de Janeiro, Brazil}
	\and
{Instituto de Astronom\'ia, Universidad Nacional Aut\'onoma de M\'exico, Ensenada, B.C., Mexico}
	   }
   \date{\today}

 
 \abstract
   {Complex bipolar shapes can be generated either as a planetary nebula or a symbiotic system. The origin of the material ionised by the white dwarf is very different in these two scenarios, and it complicates the understanding of the morphologies of planetary nebulae.}
   {The physical properties, structure, and dynamics of the bipolar nebulae, M~2--9, Mz~3, and Hen~2--104, are investigated in detail with the aim of understanding their nature, shaping mechanisms, and evolutionary history. Both a morpho-kinematic study and a spectroscopic analysis, can be used to more accurately determine the kinematics and nature of each nebula.}
   {Long-slit optical echelle spectra are used to investigate the morpho-kinematics of M~2--9, Mz~3, and Hen~2--104. The morpho-kinematic modelling software SHAPE is used to constrain both the morphology and kinematics of each nebula by means of detailed 3-D models. Near-infrared (NIR) data, as well as optical, spectra are used to separate Galactic symbiotic-type nebulae from genuine planetary nebulae by means of a 2MASS \textit{J--H/H--Ks} diagram and a $\lambda$4363/H$\gamma$ vs. $\lambda$5007/H$\beta$ diagnostic diagram, respectively.}
   {The best-fitted 3-D models for M~2--9, Mz~3, and Hen~2--104 provide invaluable kinematical information on the expansion velocity of its nebular components by means of synthetic spectra. The observed spectra match up very well with the synthetic spectra for each model, thus showing that each model is tightly constrained both morphologically and kinematically. Kinematical ages of the different structures of M~2--9 and Mz~3 have also been determined. Both diagnostic diagrams show M~2--9 and Hen~2--104  to fall well within the category of having a symbiotic source, whereas Mz~3 borders the region of symbiotic and young planetary nebulae in the optical diagram but is located firmly in the symbiotic region of the NIR colour-colour diagram. The optical diagnostic diagram is shown to successfully separate the two types of nebulae, however, the NIR colour-colour diagram is not as accurate in separating these objects.}
  {The morphology, kinematics, and evolutionary history of M~2--9, Mz~3, and Hen~2--104 are better understood using the interactive 3-D modelling tool {{\sc shape}}. The expansion velocities of the components for each nebula are better constrained and fitted with a vector field to reveal their direction of motion. The optical and NIR diagnostic diagrams used are important techniques for separating Galactic symbiotic-type nebulae from genuine planetary nebulae.}


\authorrunning{Clyne et al}
   \maketitle
%

\section{Introduction}

The formation of complex bipolar optical nebulae, such as M~2--9, Mz~3, and Hen~2--104, continue to intrigue and perplex astronomers. There are competing theories to explain the bipolar morphologies observed: common envelope evolution \citep{Nordhaus2006} or intrinsically bipolar ejections \citep{Balick2002}. Other features, such as knots or clumps, are most often associated with interactions between slow-moving ejecta and a later fast wind \citep{Sadakane2012}.  

A planetary nebula can form when a low mass star evolves through the AGB phase and subsequent thermal pulses, exposing a C-O white dwarf (WD) core with an effective temperature high enough that Lyman continuum photons ionise the ejected envelope. Planetary nebulae (PNe) are observed to exhibit a wide variety of morphologies and in general a binary system is not a prerequiste to forming a PNe.  However, for strongly bipolar planetary nebulae (bPNe), often with collimated outflows and point symmetric structures a binary central system is widely believed to be essential for the extreme deviation from spherical symmetry seen.  

In a binary system of two low mass stars, when the original PN from the higher mass primary star has dispersed, the secondary star may then in turn evolve away from the main sequence and form a second planetary nebulae. For widely spaced binary stars, the second PN may appear a little different to that from a single star. A symbiotic binary system (hereafter SySt), however, is a long-period interacting binary system composed generally of a hot white dwarf and an evolved mass-losing cold red-giant star, usually a Mira variable. In SySts, mass that is transferred and ejected from the secondary can be ionised by the old primary WD, forming another, different kind of PN termed a symbiotic nebula that shares many of the overall features of a classical PN. 

PNe and SySts are at distinctly different stages of stellar evolution. Since a SySt consists of a mass-losing Mira (or a red giant that has not yet reached the PN stage) and a white dwarf component that has evolved past the PN stage, they can be described as being simultaneously post-PN and pre-PN \citep{Lopez2004}. While SySts are binaries by definition, PNe may occur from either single- or multiple-star systems. It is not obvious whether a given bipolar nebula is a classical planetary nebula or a symbiotic nebula. Symbiotic nebulae are sometimes misclassified as PNe, since both types of objects share the same characteristics (i.e., morphology and ionization structure). Additional evidence is required to identify a SySt. 

The optical spectra from symbiotic nebulae show both absorption and emission lines from the surface of the cool red giant and the surrounding nebula, respectively. This requires the presence of a hot compact source of ionising radiation. Since both PNe and symbiotics differ vastly in their density, the change in intensities of their {[O~{\sc iii}] emission lines at 4363~\AA~and 5007~\AA~can provide a means of separating them. {O~{\sc vi}} Raman scattered lines at 6830~\AA, 7088~\AA\ seen only in symbiotic nebulae due to the extreme high densities in their cores \citep{Schmid1989}. Broad-winged, single, or double-peak H$\alpha$ line profiles, also indicate the presence of a symbiotic core \citep{Winckel1993}. It can also be possible to directly detect the contribution of a cool giant companion in the red part of the optical regime. The continua are characterised by a dominant red continuum, typical of an F-M type giant for SySts, and a weak atomic recombination continuum in the optical for PNe. 

The NIR colour-colour diagrams \citep{Whitelock1992,Schmeja2001,Phillips2007,Santander2007} have also been used to separate symbiotic nebulae from genuine PNe. On a \textit{J--H/H--Ks} colour-colour diagram, the location of symbiotic nebulae can be found quite far from those of PNe and lie in the region of increasing \textit{J--H} and \textit{H--Ks}, as expected for D-type symbiotic nebulae \cite[see Fig. 1 from][]{Phillips2007}. Nevertheless, the locus of symbiotic nebulae are also mixed with other dusty celestial objects such as Be, B[e], WR stars, or H {\sc ii} regions, that make their classification unreliable. Recently, an additional diagnostic diagram was proposed as a way to distingush symbiotic nebulae from genuine PNe by Corradi and collaborators, based on their r\textquotesingle--H$\alpha$ and r\textquotesingle--i\textquotesingle~colours \citep{Corradi2008,Corradi2010,Rodriguez2014}.

Interacting binaries have been shown to produce complex environments, which makes them very useful for studying the late stages of stellar evolution, as well as the mass loss exchange between the components. It is generally accepted that accretion disks are needed to launch jets and possibly ansae \citep{Goncalves2001,Crocker2002,Corradi2003,Sokoloski2003,Corradi2011,Werner2014}. 

To understand the origin of the range of morphologies of PNe and SySts requires both being able to correctly identify the origin of the ionised material and fully characterising the morphology and kinematics of the nebula. Here, two well known extreme bPNe (M~2--9 and Mz~3) and one confirmed symbiotic nebula (Hen~2--104) are investigated. It has been proposed that a SySt mechanism has led to the formation of M~2--9 \citep{Doyle2000,Livio2001,Schmeja2001,Smith2005,Corradi2011}, Mz~3 \citep{Schmeja2001,Guerrero2004,Smith2005}, and Hen~2--104 \citep{Whitelock1987,Lutz1989,Schwarz1989,Corradi2001,Santander2008}. 

Previous studies on the morpho-kinematics of the structures of M~2--9 and Mz~3 have revealed similar morphological features (i.e., bipolar lobes), as well as the presence of material expanding from their central core up to a few hundred km s$^{-1}$ \citep{Lopez1983,Balick1989,Solf2000,Guerrero2002}.
These materials have been proposed to be associated with a highly expanding stellar wind from a cold companion. \cite{Lykou2011} and \cite{Castro2012} found further evidence of a dusty circumstellar disk in the central region of M 2--9. This result is consistent with the presence of a close binary system but has yet to be confirmed. Regarding Hen~2--104, its symbiotic nature has been confirmed by \cite{Whitelock1992} after detecting a Mira companion in its central zone. The equatorial expansion velocity of this nebula, which has a value of 12 km s$^{-1}$ \citep{Santander2008}, is significantly lower than those found in M~2--9 and Mz~3. 

We use optical and infrared diagnostic diagrams such as a $\lambda$4363/H$\gamma$ vs. $\lambda$5007/H$\beta$ and a 2MASS NIR colour-colour diagram to demonstrate that these methods discriminate well between single ionising stars and symbiotic central stars. This technique \citep{Gutierrez1995} demonstrates that large numbers of PNe and symbiotics (D-- and S--types) can be readily separated. Thus, using these diagnostic diagrams, we characterise the central sources of M~2--9, Mz~3, and Hen~2--104. We then use the 3-D morpho-kinematic code called {\sc shape}\footnote{{\sc shape} is accessible online at http://www.astrosen.unam.mx/shape/} \citep{Steffen2011} to carefully reproduce the structural and spectral features of each nebula in three dimensions. For two of the three nebulae, it appears highly likely that a symbiotic central system is present and in all three, a complex multicomponent velocity structure is present within the overall bipolar morphologies. 

This paper is organised as follows: archival and source data and their use is presented in Sect.~\ref{data}. Analyses and results of each topic are presented in Sect.~\ref{analysis}. This section includes: the modelling of optical kinematic data for the objects M~2--9, Mz~3, and Hen~2--104, using {\sc shape}, and a spectroscopic analysis of M~2--9, Mz~3, Hen~2--104, and a number of other symbiotic and PNe sources in the optical and infrared using a $\lambda$4363/H$\gamma$ vs. $\lambda$5007/H$\beta$ diagnostic diagram and a 2MASS \textit{J--H/H--Ks} diagram. Discussion and conclusions are found in Sects.~\ref{discussion} and \ref{conclusions}, respectively.


\section{Archival data}
\label{data}

\subsection{Long-slit echelle spectra}

The observed long-slit echelle spectra used to assist with the creation of our model of M~2--9 were obtained from the San Pedro M\'artir kinematic catalogue of Galactic Planetary Nebulae\footnote{http://kincatpn.astrosen.unam.mx} \citep{Lopez2012}. Their observations were taken on June 2004 at the 2.1 m telescope in San Pedro M\'artir National Observatory, Mexico, with the Manchester Echelle Spectrometer \citep[MES;][]{Meaburn2003}. Each slit had a slit width of 0.9$^{\prime\prime}$ and a velocity resolution of 5.75 km s$^{-1}$. The spatial scale and sampling used for each slit were 0.312$^{\prime\prime}$ pixel$^{-1}$ and 2.3 km s$^{-1}$ pixel$^{-1}$, respectively. The spectra in {[N~{\sc ii]} were chosen for the morpho-kinematic modelling of M~2--9 because of the lower thermal broadening compared to the H$\alpha$ line (thermal width of H$\alpha$ and {[N~{\sc ii]} at 10,000 K are 21.4 km s$^{-1}$ and 5.8 km s$^{-1}$, respectively).

The observed long-slit echelle spectra used to assist with the creation of our model of Mz~3 were obtained from \cite{Guerrero2004}. Their observations were taken on June 2002 using the echelle spectrograph on the CTIO 4 m telescope at Cerro Tololo, Chile. A slit width of 0.9$^{\prime\prime}$ and a velocity resolution of 8 km s$^{-1}$ were used for each slit position. The spatial scale and sampling used were 0.26$^{\prime\prime}$ pixel$^{-1}$ and 3.7 km s$^{-1}$ pixel$^{-1}$, respectively. Optical seeing was better than 1.2$^{\prime\prime}$ during the observed period.

The observed long-slit echelle spectra used to assist with the creation of our model of Hen~2-104 were obtained from \cite{Corradi2001}. Their observations were taken on June 2000 with the CTIO 4 m telescope. The slit width used was 1.5$^{\prime\prime}$ and the spatial scale along the slit was 0.54$^{\prime\prime}$ pixel$^{-1}$. The sampling and resolution used were 3.7 km s$^{-1}$ pixel$^{-1}$ and 13 km s$^{-1}$, respectively. The seeing varied between 1.5$^{\prime\prime}$ and 1.9$^{\prime\prime}$ during the observations.

\subsection{Archival data for diagnostic diagrams}

The emission lines used for the objects in the optical diagnostic diagram and hence the separation between SySts and PNe were obtained from various works in the literature 
(see Tables~\ref{table:optss} and~\ref{table:optpne} in the appendix). The 2MASS colour magnitudes of the same sample of SySts and PNe were obtained from the All--Sky Catalog of Point Sources NASA/IPAC Infrared Science Archive: \citet{Cutri2003b,Cutri2003}. All colour magnitude values are summarised in Tables~\ref{table:2massnir} and \ref{table:2masspne}.


\section{Analyses and results}
\label{analysis}

\subsection{Morpho-kinematic modelling of M~2--9}
\label{m2-9} 

The Galactic PN M~2--9, also called Minkowski's Butterfly Nebula or PN G010.8+18.0 \citep{Acker1992}, is an enigmatic object with intense aspherical flows and displays a highly collimated bipolar structure \cite[120$^{\prime\prime}$$\,\times\,$12$^{\prime\prime}$,][]{Schwarz1997}. Its bipolar structure contains two separate shells, an inner bulb and an outer neutral shell \citep{Corradi2011}, which are thought to be formed from two equatorial rings/tori \citep{Castro2012} found at the nebula's centre. The nebula also has a feathered component which extends all the way out to a pair of dusty blobs. Sub-structures in the form of arcs and surface brightened spots are also apparent.

Apart from the dusty blobs and extended material (matter in the region beyond which the feathers disappear in the [N~{\sc ii}] image), we have managed to construct and constrain the morpho-kinematics of M~2--9 and its components using a 3-D astrophysical modelling tool called {\sc shape} (version 5.0). One aim involved creating the 3-D morphology resembling the nebula and its components using observed spectra (optical long-slit echelle spectra), which are compared and matched up with synthetic spectra generated from our model. It should be noted that the same procedure described is also followed for the construction of Mz~3 and Hen~2--104.

%
\begin{figure*}
\centering
\includegraphics[width=16.2cm]{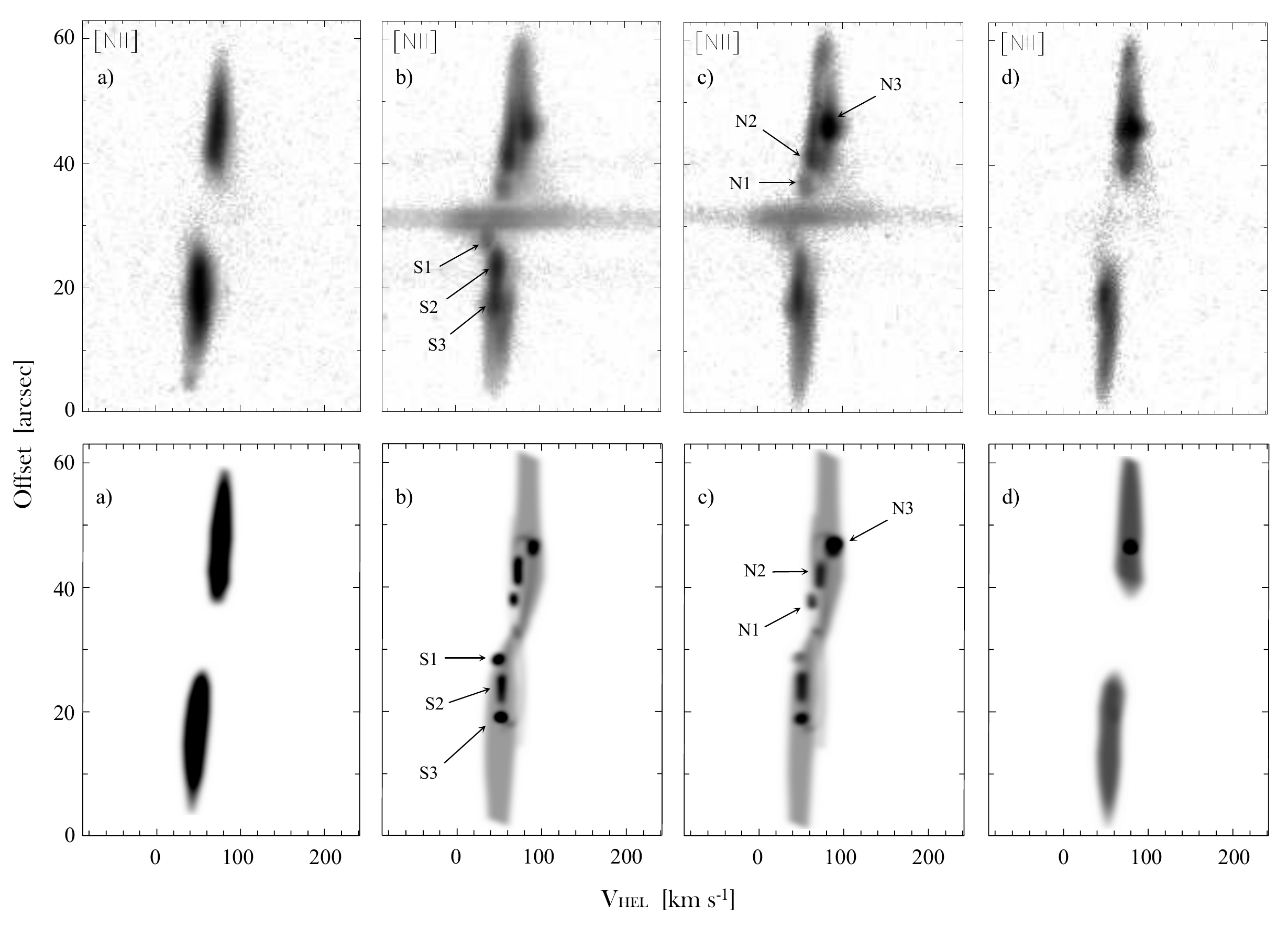}
\caption{Observed [N~{\sc ii}] $\lambda$6583 (top row) and synthetic (bottom row) spectra for M~2--9. The brightness enhanced features are labelled on both sets of spectra for clarity. The systemic velocity $V_{\rm sys}$ = 69 km s$^{-1}$. The observed spectra shown are from \cite{Lopez2012}.}
\label{fig:fig4}
\end{figure*}
%

%
\begin{figure*}
\centering
\includegraphics[width=13.8cm]{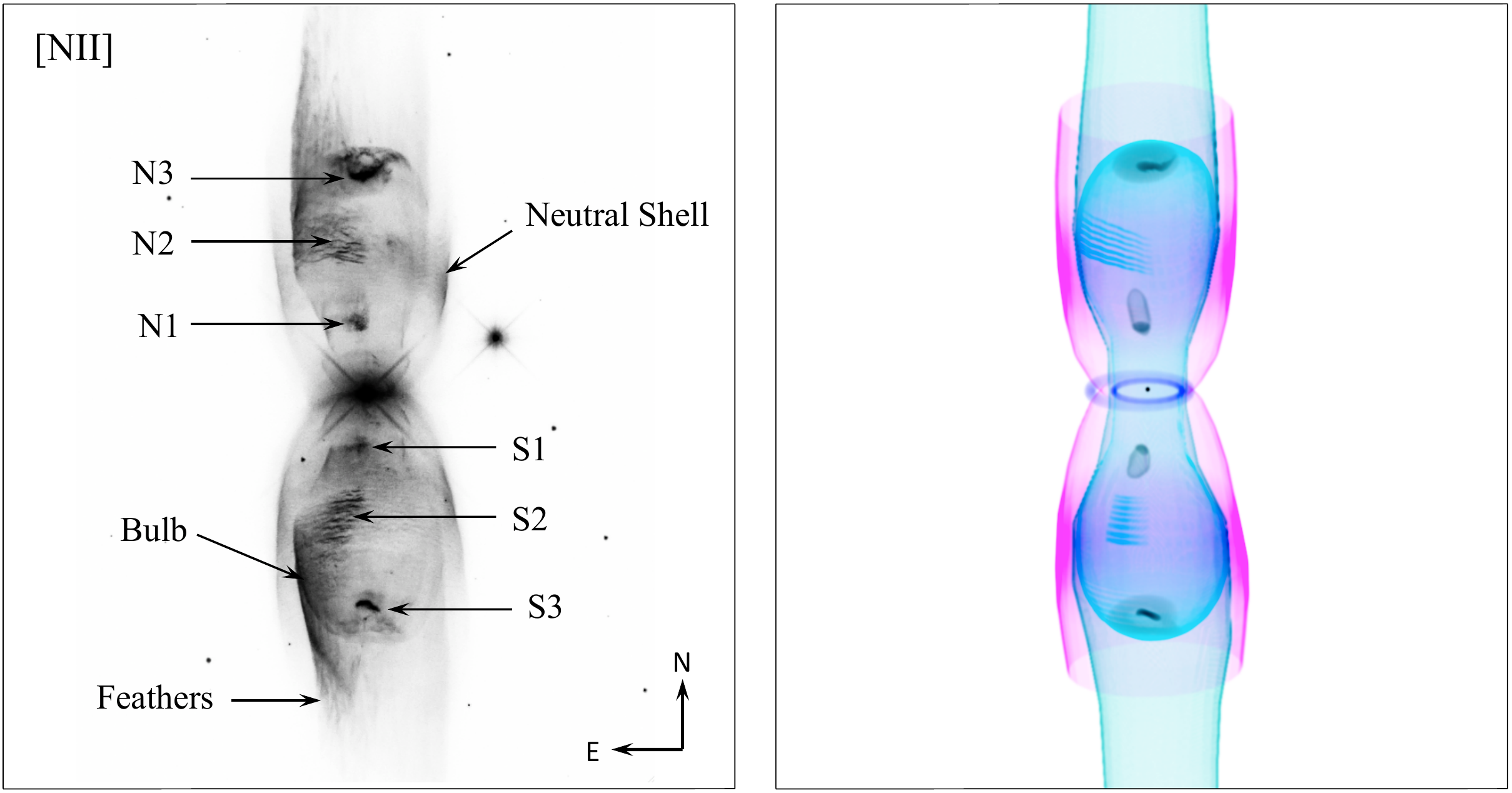}
\caption{Grayscale \textit{HST} [N~{\sc ii}] image (left) and {\sc shape} model (right) of M~2--9. The main components and features of the nebula are labelled in the [N~{\sc ii}] image. Both equatorial rings seen across the minor axis in the {\sc shape} image are best observed in the $^{12}$CO~$\textit{j}=2-1$ line emission \citep{Castro2012}. The height and width of each box is 48$^{\prime\prime}$x 48$^{\prime\prime}$.}
\label{fig:m29image}
\end{figure*}
%

Development of the model involved the use of five individual components; an inner closed-end bipolar component (\textit{bulb}), a surrounding feathered shell (\textit{feathers}), an outer bipolar component (\textit{neutral shell}), and two tori (inner torus \textit{ring 1} and outer torus \textit{ring 2}). The arcs (\textit{N2,S2}) and surface brightened regions (\textit{N1,N3,S1,S3}) where created by brightness enhancing of the bulb component surface. The names of the components were adopted from \cite{Kohoutek1980}, \cite{Corradi2011}, and \cite{Castro2012}. Each individual component was modified using modifiers such as; \textit{density}, \textit{velocity}, \textit{shell}, \textit{squeeze}, \textit{rotation}, etc. For example, the \textit{density} modifier sets the spatial distribution of density, the \textit{velocity} modifier applies a velocity field as a function of position, the \textit{shell} modifier creates a shell of given thickness, the \textit{squeeze} modifier compresses or expands an object as a function of position along the local symmetry axis of the modifier, and the \textit{rotation} modifier allows a chosen object (sphere, torus, etc,.) to be rotated about a specific point.} A number of density and squeeze modifiers were customised interactively instead of analytically to carefully produce accurate results.

%
\begin{table}[ht]
\caption{\label{table:para}{\sc shape} components and velocities of M~2--9.}
\centering
\begin{tabular}{lccl}
\hline\hline\addlinespace[3pt]
Component & location & & velocity (km s$^{-1}$) \\ \addlinespace[1pt]
\hline \addlinespace[3pt]
Bulb & northern pole & & $96\pm3$ ($27\pm3$) \\ \addlinespace[1pt]
 & southern pole & & $40\pm3$ ($-29\pm3$) \\ \addlinespace[1pt]
Feathers & north$_{\rm max}$ & & $85\pm3$ ($16\pm3$) \\ \addlinespace[1pt]
 & south$_{\rm max}$ & & $39\pm3$ ($-30\pm3$) \\ \addlinespace[1pt]
Neutral Shell & north$_{\rm max}$ & & $97\pm2$ ($28\pm2$) \\ \addlinespace[1pt]
& south$_{\rm max}$ & & $41\pm2$ ($-28\pm2$) \\ \addlinespace[1pt]
$^{\dagger}$Ring 1 & ---- & & $72.9\pm0.1$ ($3.9\pm0.1$) \\ \addlinespace[1pt]
$^{\dagger}$Ring 2 & ---- & & $76.8\pm0.1$ ($7.8\pm0.1$) \\ \addlinespace[2pt]
\hline
\end{tabular} 
\tablefoot{The velocities shown are values with respect to $V_{\rm sys}$ = +69 km s$^{-1}$. The values in parentheses are given as $V-V_{\rm sys}$ (velocities corrected for $V_{\rm sys}$). Here north$_{\rm max}$ and south$_{\rm max}$ refers to the material found furthest out since these type of objects do not have a peak polar region. $^{\dagger}$velocities adopted from \cite{Castro2012}.}
\end{table}
%

In order to mimic the observed spectra and constrain the kinematics of M~2--9, each structure and sub-structure required a velocity component. The velocity function in {\sc shape} is given by $|$\,f\,$|$\,(r) = A + B(\textit{r/r$_{0}$}) + C(\textit{r/r$_{0}$})$^{D}$, where \textit{r} is the initial arbitrary (constant) radius given to the structure, whereas all the other parameters are variables. The \textit{r$_{0}$} parameter is used to give the rate of change of velocity with respect to the radius \textit{r}. Homologous and radial vector fields were also assigned to the different components to account for their type of expansion. There is some evidence however of localised deviations from a radial flow in the outer lobes of M~2--9, where the addition of a modest laterally expanding velocity component provides a better fit. See Table.~\ref{table:para} for the velocities of each structure and the appendix for the 3-D mesh images of M~2--9.

The four long-slit spectra of \cite{Lopez2012} were used to create and test our model of M~2--9. All four slits are positioned parallel to the nebula's symmetry axis (north to south). Slit {\it c} is placed directly down the symmetry axis, whereas slit {\it a}, {\it b}, and {\it d} are offset from the axis by $-5.0^{\prime\prime}$, $-3.0^{\prime\prime}$, and $+3.5^{\prime\prime}$, respectively. It should be noted that the observed long-slit spectra from \cite{Solf2000} and \cite{Smith2005} were also used to help assist with the modelling. A slit width and slit length of 1$^{\prime\prime}$ and 60$^{\prime\prime}$ respectively were used to render the synthetic spectra. In each case, the synthetic spectra are rendered at the same velocity range (330 km s$^{-1}$) and spectral resolution (5.75 km s$^{-1}$) as the observed spectra. We should also note that the velocities used for the modelling of ring 1 and ring 2 were adopted from \cite{Castro2012} due to the intensity of [N~{\sc ii}] emission from the central region (see observed spectra \textit{b} and \textit{c} in Fig.~\ref{fig:fig4}). The systemic velocity $V_{\rm sys}$ for the observed and synthetic spectra is +69 km s$^{-1}$ \citep[see][]{Smith2005}. Both sets of spectra are presented alongside each other in Fig.~\ref{fig:fig4}. The \textit{HST} difference image\footnote{http://apod.nasa.gov/apod/ap130915.html} and rendered {\sc shape} image can be seen in Fig.~\ref{fig:m29image}.

Our best-fit model (including all bipolar components) has an inclination of its symmetry axis with respect to the plane of the sky of $i = 17^{\circ}$, which agrees with the value found by \cite{Zweigle1997} and \cite{Solf2000}, but is slightly smaller than the value ($i = 18^{\circ}$) found by \cite{Corradi2011}. A marginally smaller value of $i = 16^{\circ}$ was found by \cite{Lykou2011}, whereas values of $i = 15^{\circ}$ \citep{Schwarz1997} and $i = 11^{\circ}$ \citep{Goodrich1991} were also determined.

The distance to M~2--9 has been routinely quoted throughout the literature and is said to have values of 650~pc \citep{Schwarz1997,Doyle2000,Solf2000,Castro2012}, 1.2~kpc \citep{Lykou2011}, and 1.3~kpc \citep{Corradi2011}. In this study of M~2--9 we adopt the distance of 650~pc. At this distance we find a kinematical age of $915\pm90$ yr for the bulb. \cite{Castro2012} found a similar age for the inner molecular torus (ring 1) of $\sim$900 yr at 650 pc, which suggests that the bulb may have formed just after the formation of ring 1. The kinematical age we find for the neutral shell is $1430\pm90$ yr, whereas \cite{Solf2000} determined an age of 1300 yr. \cite{Castro2012} and \cite{Zweigle1997} found values of 1400 yr and 1365 yr for ring 2, respectively, which suggests, like that of the bulb, that the neutral shell could have formed after the formation of ring 2. This evidently indicates that both rings are the equatorial counterparts of these bipolar outflows. 

Regarding the feathered component, we were unable to determine an accurate kinematical age because our modelled spectra does not include the extended material and dusty blobs (N4 and S4) which most likely makes part of the feathers. \cite{Corradi2011} found the kinematical age of these blobs to be 2500 yr for a distance of 1.3~kpc and a velocity of 147~km~s$^{-1}$. However, considering a distance of 650~pc with the same velocity of 147~km~s$^{-1}$, the kinematical age of the blobs would be $\sim$1250 yr. This indicates that the neutral shell was the earliest ejection event, then followed by the feathers and dusty blobs, and most recently the bulb.  

\subsection{Morpho-kinematic modelling of Mz~3}
\label{mz3}

Menzel 3 (PN G331.7--01.0), or Mz~3 as it is commonly known, is an intrinsically bright young bipolar PN, which has arguably the most complex bipolar morphology seen to date. Mz~3 consists of three nested pairs of bipolar lobes (\textit{BL1}, \textit{BL2}, and \textit{BL3}), an equatorial ellipse-like feature (\textit{EE}), and three knots (\textit{knot 1}, \textit{knot 2}, and \textit{knot 3}), which we adopted from \cite[hereafter Gu04]{Guerrero2004}. BL1 appears to have a closed-lobe morphology, whereas BL2 and BL3 have open lobes with cylindrical and conical shapes, respectively. The faint equatorial ellipse of emission appears to be aligned along the equator of the nested bipolar lobes (see Fig. 1 of Gu04 for a more noticeable view of this component). The three faint knots, two (knot 1 and knot 2) of which are located just above the apex of the northern lobe of BL1, and one (knot 3) found just below the southern lobe, were initially discovered by Gu04.

%
\begin{figure*}
\centering
\includegraphics[width=14cm]{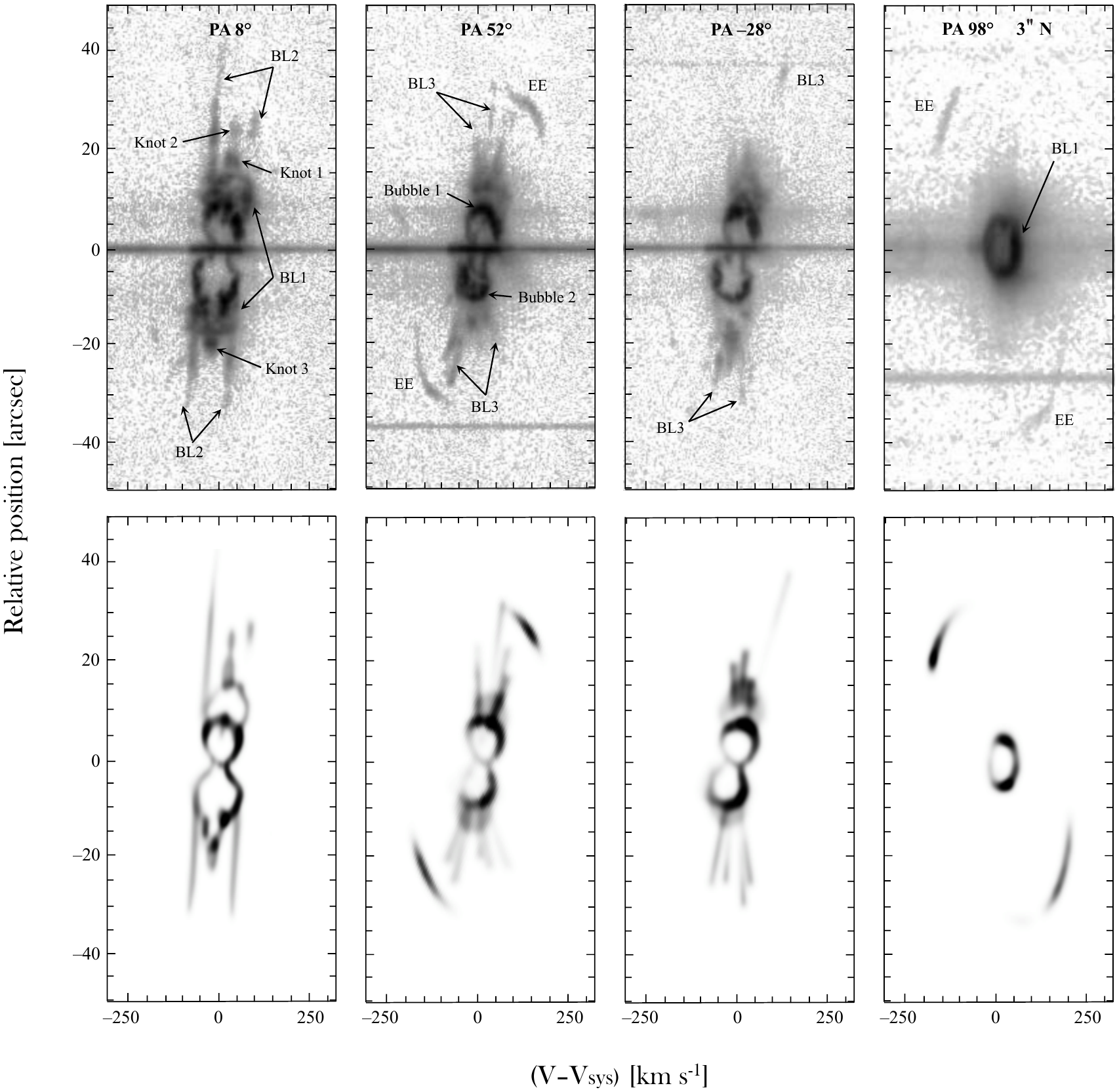}
\caption{Observed [N~{\sc ii}] $\lambda$6583 (top row) and synthetic (bottom row) spectra for Mz~3. The P.A.s and nebular components are labelled on the observed spectra. The velocities are corrected for the systemic velocity of the nebula. The observed spectra shown are from \cite{Guerrero2004}.}
\label{fig:fig5}
\end{figure*}
%

%
\begin{figure*}
\centering
\includegraphics[width=14.35cm]{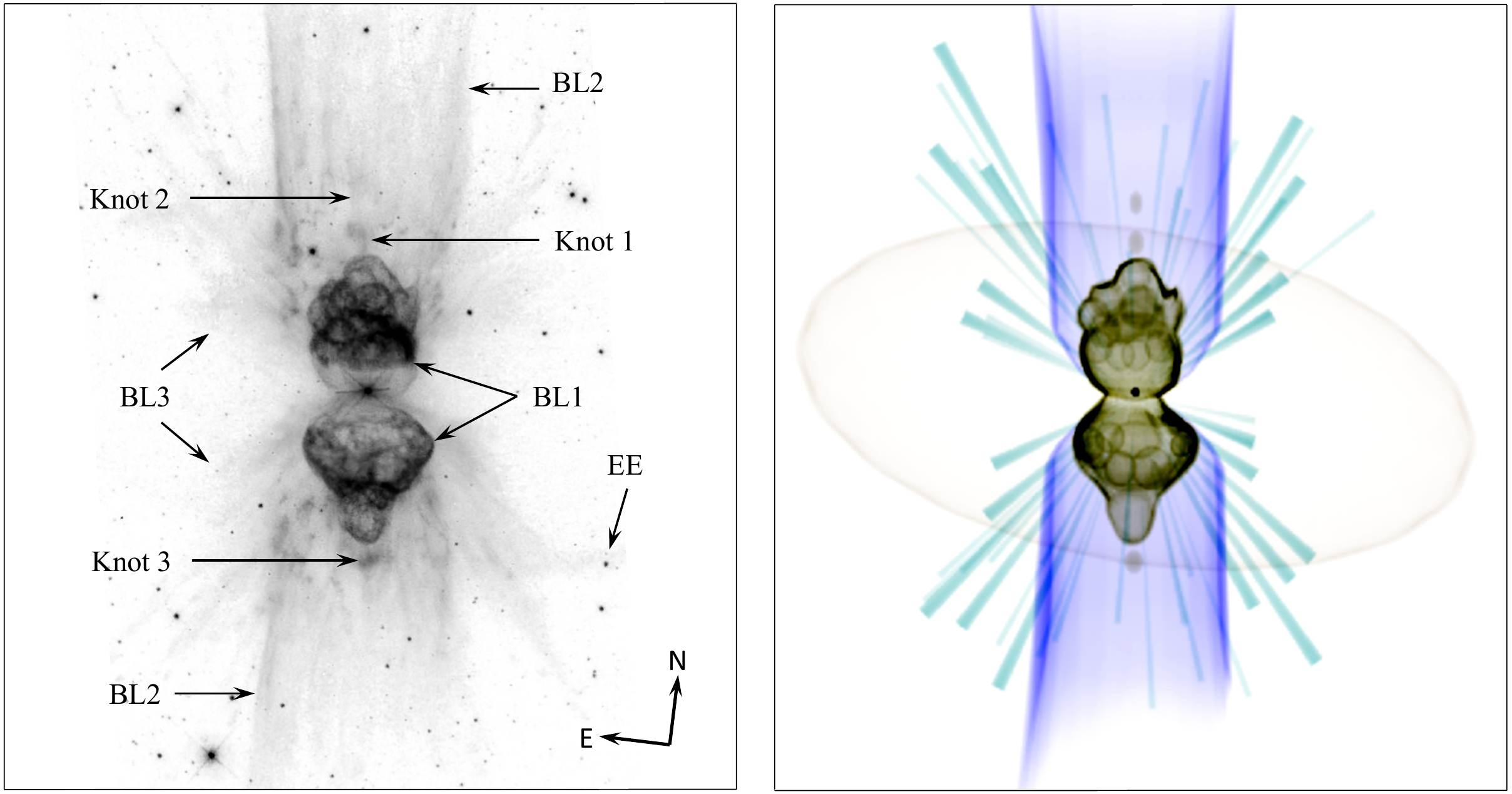}
\caption{Grayscale composite \textit{HST} image (left) and {\sc shape} model (right) of Mz~3. The main components of the nebula are shown in the \textit{HST} image. The equatorial ellipse (EE) is difficult to observe in the composite \textit{HST} image but is best seen in Fig.~1 of Gu04. The height and width of each box is 96$^{\prime\prime}$x 96$^{\prime\prime}$.}
\label{fig:mz3image}
\end{figure*}
%

To get a better understanding of Mz~3's complex morphology and kinematics, a {\sc shape} model of the nebula was created. Completion of this model included seven components, two of which contained a number of sub-components. For example, the inner bilobed structure BL1, or \textit{IBL} in \cite{Redman2000}, consists of two inner bubbles (\textit{Bubble 1} and \textit{Bubble 2}) that represent what appears to be a pair of inner bilobes, and the top region of each lobe of BL1 contains twelve sub-bubbles that represent the foamy texture, or blister-like structures in Gu04, whereas the outer bipolar conical structure BL3, or \textit{rays} in \cite[hereafter SG04]{Santander2004}, is made up of fifty-eight individual ray-like structures. See Table~\ref{table:paramz3} for all components and velocities used to create our model of Mz~3.

The observed spectra (Fig.~\ref{fig:fig5}) used to assist in the creation of Mz~3 are those from Gu04. Their high-dispersion spectroscopic observations allowed us to determine kinematic properties of the different components, as well as tightly constrain the nebula's morphology. The four slit positions at P.A.s $-28^{\circ}$, 8$^{\circ}$, 52$^{\circ}$, and 98$^{\circ}$ were considered because of the important regions they cover on the nebula (see Fig. 3 from Gu04 for a detailed image of these slit positions). The different morphological components of Mz~3 are labelled on the observed spectra, as clearly shown in Fig.~\ref{fig:fig5}. The height and velocity range for both sets of spectra are 100 arcsec and 600 km s$^{-1}$, respectively. The difference image\footnote{http://www.astro.washington.edu/users/balick/WFPC2/index.html}, which is a composite grayscale \textit{HST} image, and the rendered {\sc shape} image are shown in Fig.~\ref{fig:mz3image}. See appendix for the 3-D mesh images of Mz~3.

%
\begin{table}
\caption{\label{table:paramz3}{\sc shape} components and velocities of Mz~3.}
\centering
\begin{tabular}{lcc}
\hline\hline\addlinespace[3pt]
Component & location & velocity (km s$^{-1}$) \\ \addlinespace[1pt]
\hline \addlinespace[3pt]
BL1 & northern pole & \hspace{0.2cm}$100\pm5$ \\ \addlinespace[1pt]
 & southern pole & \hspace{0.15cm}$-80\pm5$ \\ \addlinespace[1pt]
BL2 & north$_{\rm max}$ & \hspace{0.2cm}$110\pm7$ \\ \addlinespace[1pt]
 & south$_{\rm max}$ & $-100\pm7$ \\ \addlinespace[1pt]
BL3 & north$_{\rm max}$ & \hspace{0.4cm}$140\pm10$ \\ \addlinespace[1pt]
& south$_{\rm max}$ & \hspace{0.2cm}$-100\pm10$ \\ \addlinespace[1pt]
EE & equatorial & \hspace{0.4cm}$200\pm10$ \\ \addlinespace[1pt]
Knot 1 & ---- & \hspace{0.4cm}$30\pm2$ \\ \addlinespace[1pt]
Knot 2 & ---- & \hspace{0.4cm}$40\pm2$ \\ \addlinespace[1pt]
Knot 3 & ---- & \hspace{0.2cm}$-30\pm2$ \\ \addlinespace[2pt]
\hline
\end{tabular} 
\tablefoot{The velocities shown are given as $V-V_{\rm sys}$.}
\end{table}
%

All three pairs of bipolar lobes share the same axis of symmetry (Gu04, SG04) and each one is modelled with an inclination of its symmetry axis with respect to the plane of the sky of $i = 12^{\circ}$. An inclination between 12$^{\circ}$ and 20$^{\circ}$ was found by Gu04, whereas SG04 find a value between 14$^{\circ}$ and 19$^{\circ}$, and \cite{Lykou2007} proposed a value of $i = 16^{\circ}$.

The distance to Mz~3 remains uncertain and a range of values have been previously published; 1~kpc \citep{Lopez1983}, 1.3~kpc \citep{Cahn1992,Pottasch2005}, 1.4~kpc \citep{Lykou2007}, 1.8~kpc \citep{Cohen1978}, and 2.7~kpc \citep{Kingsburgh1992}. Due to this uncertainty, we will show explicitly the dependence of the kinematical age on distance, like those shown in Gu04 and SG04.

SG04 show BL1 to have a kinematical age of $tD^{-1}$ (yr kpc$^{-1}$) = 670 yr with a possible range of 550--710 yr, whereas Gu04 find BL1 to be $600\pm50(D$/kpc) yr, where $D$ is the distance in kiloparsecs. We find a kinematical age of $650\pm35(D$/kpc) yr. Also, the inner bipolar lobes (bubble 1 and bubble 2) found within BL1 interestingly share the same kinematical age as BL1 for a $v_{\rm max}$ = 65~km~s$^{-1}$ and $\sim$9$^{\prime\prime}$ displacement from the nebular centre. This may suggest that the inner bipolar lobes and BL1 most likely formed at the same time.     

The knots, found just above the poles of BL1, were modelled with displacements (values are from Gu04) of 19$^{\prime\prime}$ (knot 1), 24$^{\prime\prime}$ (knot 2), and 19$^{\prime\prime}$ (knot 3) from the central star and along the symmetry axis. The knots are shown to have kinematical ages of $1020\pm70(D$/kpc) yr (knot 1), $1200\pm60(D$/kpc) yr (knot 2), and $1020\pm70(D$/kpc) yr (knot 3), taking into consideration that they lie along the symmetry axis of BL1 and follow that of a Hubble flow. 

The morphology of BL2, or \textit{columns} in SG04, is quite different to that of BL1. It has a highly collimated cylindrical shell, which increases gradually in diameter {\bf ($d$)} along each lobe from the end point of the curvature ($d\sim$20$^{\prime\prime}$) near its centre to its furthest observable point out ($d\sim$30$^{\prime\prime}$). BL2 also has a tight waist (2.8$\pm$0.2$^{\prime\prime}$) and a relatively thick shell (2.5$\pm$0.2$^{\prime\prime}$). We find a kinematical age for BL2 to be $1000\pm50(D$/kpc) yr, which confirms the same age found by Gu04. A value of 875 yr, with a possible range 830--1000 yr , was acquired by SG04 for BL2.

A more complicated structure to create was that of BL3. Although it is a single structure by nature, here it is made up of fifty-eight individual rays to give its stringy-like appearance. Each ray was created as a long thin cone to give the effect of expansion of the material. These rays are positioned in such a way that they give a conical bipolar shape to the entire structure. The opening angles for the lobes of BL3 are $\sim$55$^{\circ}$ (47$^{\circ}$ and 50$^{\circ}$ in SG04 and Gu04 respectively) with respect to the axis of symmetry, and are positioned around a 360$^{\circ}$ angle with the nebular centre as the point of emanation. For a maximum extension of the rays $\sim$55$^{\prime\prime}$ (60$^{\prime\prime}$ in Gu04) and adopting a Hubble-like flow, we determine a kinematical age of $1800\pm100(D$/kpc) yr, which agrees with the value found in Gu04. A value of 1600 yr, with a possible range 1400--1800 yr, was obtained by SG04 for BL3.

On the other hand, the EE component was modelled as an elliptical shell with a higher expansion velocity at its equatorial regions than at its poles. Like that of the three bipolar components, we modelled EE with an inclination of 12$^{\circ}$ but with its symmetry axis (that connecting its poles) tilted by +8$^{\circ}$ (P.A. 0$^{\circ}$). Using our modelled parameters, we find a kinematical age of $950\pm50(D$/kpc) yr, which is in close agreement with the value found by Gu04 ($1000\pm50(D$/kpc) yr).
Therefore, considering the kinematical ages derived, and those compared with Gu04 and SG04, the rays of BL3 would be first produced, then followed by the combination of BL2, EE and the knots, and most recently the BL1 structure.

\subsection{Morpho-kinematic modelling of Hen~2--104}
\label{hen2-104}

The Southern Crab (PN G315.4+09.4), or Hen~2--104, is a symbiotic nebula that exhibits an ionised bipolar outflow as a result of a symbiotic binary system at its centre. Hen~2--104 was once classified as a PN until \cite{Whitelock1987} recognised a long-period Mira at its centre. The formation of bipolar nebulae around symbiotic stars are said to be the result of complex interactions, and Hen~2--104 clearly appears to be one such example. 

%
\begin{figure*}[!]
\centering
\includegraphics[width=17.4cm]{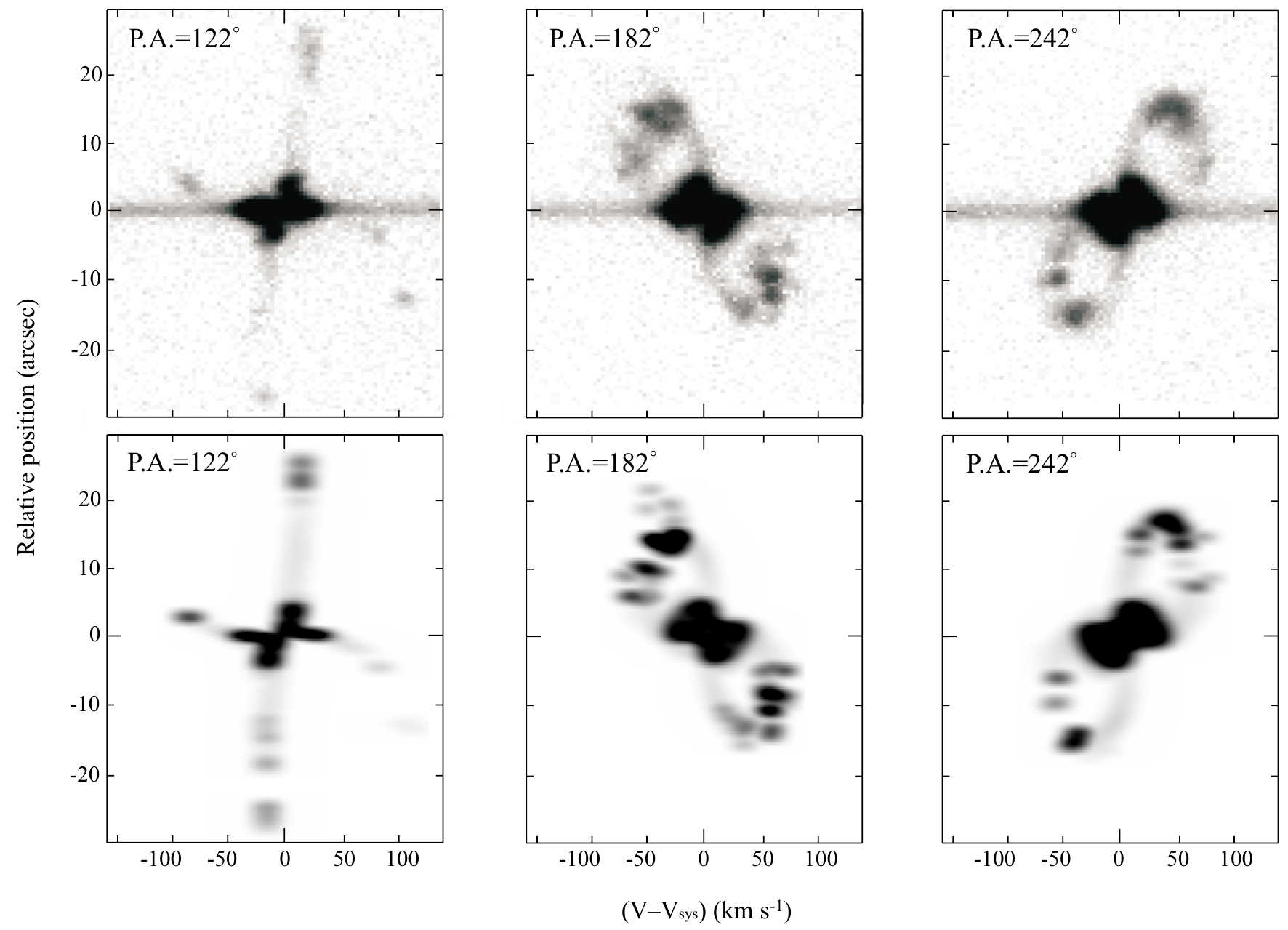}
\caption{Observed [N~{\sc ii}] CTIO (top row) and synthetic (bottom row) spectra for Hen~2--104. Shown are the spectra at P.A.s. 122$^{\circ}$, 182$^{\circ}$, and 242$^{\circ}$, for simplicity. The radial velocities are corrected for the systemic velocity of the nebula. The observed spectra are from \cite{Corradi2001}.}
\label{fig:henarrays}
\end{figure*}
%

%
\begin{figure*}[!]
\centering
\includegraphics[width=14cm]{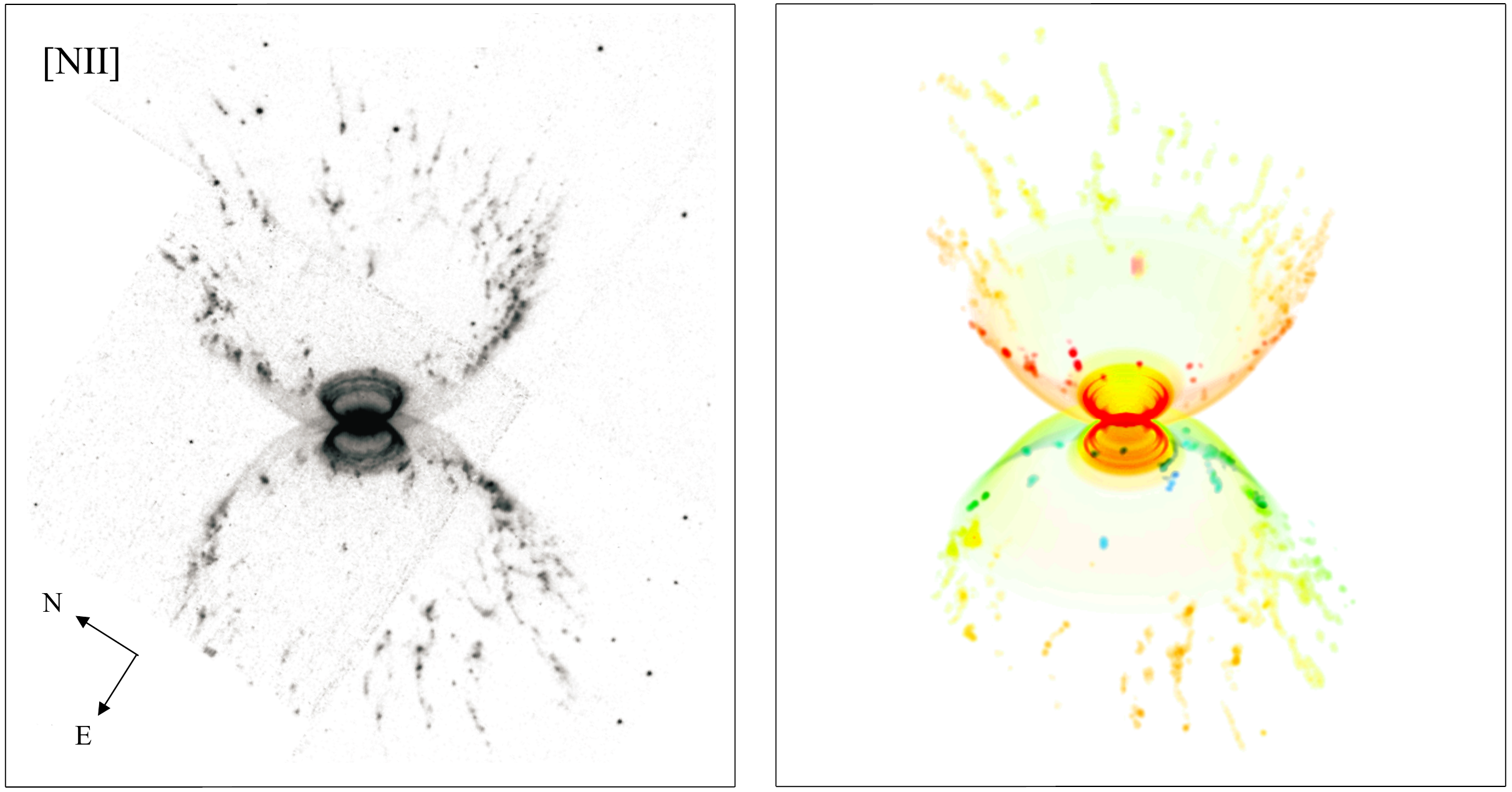}
\caption{\textit{HST} [N~{\sc ii}] image (left) and {\sc shape} model (right) of Hen~2--104. The {\sc shape} model shows the small-scale structures (fragments) of the outer shell in Doppler colour to display their direction of motion. The height and width of each box is 62.5$^{\prime\prime}$x 62.5$^{\prime\prime}$.}
\label{fig:henimage}
\end{figure*}
%

The nebula consists of three main components: an inner hourglass, an outer open-ended hourglass, and a pair of polar jets found along the symmetry axis of the system \cite[see][]{Corradi1993,Corradi2001,Santander2008}. The outer bipolar nebula has a span of 70$^{\prime\prime}$x\,35$^{\prime\prime}$ \citep{Santander2008} in the plane of the sky, and the inner hourglass is about five times smaller in scale \citep{Corradi2001}. The inner and outer hourglass structures share the same symmetry axis \citep{Corradi2001,Santander2008} and are inclined by 55$^{\circ}$ respect to the line of sight, with the southern lobe pointed towards us. \cite{Corradi2001} found an inclination $i=60^{\circ}$, whereas \cite{Santander2008} determined a value $i=58^{\circ}$. Three symmetrical rings are seen on the higher latitudes of each lobe of the inner hourglass, as shown by \cite{Corradi2001}. Where the inner hourglass appears intact, the outer hourglass is highly fragmented in the form of knots.

%
\begin{table}[ht]
\caption{\label{table:henvels}{\sc shape} components, edge distances, and constants of proportionality for the model of Hen~2--104.}
\centering
\begin{tabular}{lccc}
\hline\hline\addlinespace[3pt]
Component & edge$_{\rm inner}$ ($^{\prime\prime}$) & edge$_{\rm outer}$ ($^{\prime\prime}$) & k (km s$^{-1}$/$^{\prime\prime}$)\\ \addlinespace[1pt]
\hline \addlinespace[3pt]
Inner bipolar & ----- & 5.6 & 6 \\ \addlinespace[1pt]
Knots (red) & 10.3 & 23.5 & 2.5 \\ \addlinespace[1pt]
Knots (blue) & 12.3 & 20.0 & 2.5 \\ \addlinespace[1pt]
$^{\dagger}$Bipolar$_{\rm smooth}$ & ----- & ----- & 2.5 \\ \addlinespace[2pt]
$^{\dagger\dagger}$Jet knot$_{\rm red}$ & ----- & ----- & 10 \\ \addlinespace[2pt]
$^{\dagger\dagger}$Jet knot$_{\rm blue}$ & ----- & ----- & ----- \\ \addlinespace[2pt]
\hline
\end{tabular} 
\tablefoot{The velocity law is $v(r)$ = $k(r/r_{0})$, where $k$ (given in km~s$^{-1}$~arcsec$^{-1}$) is the constant of proportionality that maps the position to velocity, $r_{0}$ is a fiducial reference distance, and $r$ is the distance from the centre as measured in arcsecs. The deprojected distances to the edges of the components are given in arcsecs. $^{\dagger}$The outer edge of the smooth bipolar component ends where the knots begin. $^{\dagger\dagger}$The Jet knot$_{\rm red}$ has a velocity of $\sim$130 km s$^{-1}$ at 13$^{\prime\prime}$, whereas the velocity of Jet knot$_{\rm blue}$ (located at 11$^{\prime\prime}$) is unconstrained, since it is not covered by the spectroscopy.}
\end{table}
%

The overall structure of Hen~2--104 has been modelled before by \cite{Corradi2001} and \cite{Santander2008}, both of whom performed a comprehensive study of the nebula's inner and outer hourglass' and its high-velocity polar knotty outflow. However, in order to find any potential systematics in the distribution of the small-scale structure, we are interested in the full 3-D distribution of the knots and filaments. For our morpho-kinematic analysis of Hen~2--104 we used the observed long-slit spectra (Fig.~\ref{fig:henarrays}) from \cite{Corradi2001}. Six long-slit positions with P.A.s from +122$^{\circ}$ to +272$^{\circ}$, in increments of +30$^{\circ}$ and through the nebular centre, were used to help create our model of Hen~2--104. The slits positioned at P.A.s +122$^{\circ}$ and +212$^{\circ}$ pass directly through the major axis (axis of symmetry) and minor axis, respectively. The \textit{HST} difference image\footnote{http://www.spacetelescope.org/images/opo9932a/} and rendered {\sc shape} image are shown in Fig.~\ref{fig:henimage}. See Table~\ref{table:henvels} for a list of the components used to create our model of Hen~2--104.

The highly knotty and filamentary structure of Hen~2--104 required a slightly different modelling approach compared to the other objects. We therefore first produced a cylindrically symmetric mesh model of the overall structure, taking into account the \textit{HST} image and observed long-slit spectra. In particular, we considered the inner smooth low level emission and the knots and filaments further out. It turned out that the clear transition from a smooth emission distribution to the knots and filaments at an angular distance from the centre of about 17$^{\prime\prime}$ was a very useful constraint for the model. In the regions of ambiguous positioning in the image (slit positions +182$^{\circ}$ and +242$^{\circ}$), where the line of sight is nearly tangential to the lobes, we restricted the positioning of the knots to a region (outside the red and blue rings shown in the mesh images of Hen~2--104 in the appendix) where they start in the locations that are not ambiguous. Furthermore, this constraint helped to identify two inner knots near the projected axis that we now attribute to the collimated outflow along the axis, which are indicated in the bottom two mesh images as `Jet'.

To model the detailed position and structure of the individual knots we applied particles to the positions of the knots and filaments. {\sc shape} incorporates a special 3-D cursor tool whose position can be limited to a surface mesh and then located to any position with the help of the mouse or a pen device on a graphics tablet. The rendering size and brightness of the particles can also be controlled. Bright knots have been given a higher brightness than lower surface brightness filaments. This brightness assignation is not meant to reproduce the actual brightness in detail but is meant to be a rough indication of the relative brightness of the features, which is sufficient for the current purpose of our analysis.

The accuracy of the positioning of the particles in the regions where the lobe surface is seen nearly face on is close to the level of the \textit{HST} image resolution. In regions where the line of sight is nearly tangential to the lobe surface, the positioning of the knots and filaments relies mainly on the position in the lower-resolution long-slit spectra and the additional constraints mentioned above. It turns out that the positioning is still very accurate, probably only a factor of two less accurate than in the other regions.

\subsection{The optical PNe versus SySts diagram}
\label{diagnostic}

In the last twenty years, several efforts have been made to separate symbiotic-type nebulae from genuine PNe, mainly based on observations in the optical regime. One effort of this kind was the BPT diagram \citep{Baldwin1981} that attempts to separate extragalactic emission-line objects according to their excitation mechanism. Here we present a diagnostic diagram based on the 
$\lambda$4363/H$\gamma$ vs. $\lambda$5007/H$\beta$ emission line ratios for the separation of genuine PNe from SySts, as proposed by \cite{Gutierrez1995}. Fig.~\ref{fig:fig1} plots $\lambda$4363/H$\gamma$ vs. $\lambda$5007/H$\beta$ for the three objects studied here (M~2--9, Mz~3, and Hen~2--104) together with a significant sample of known SySts and PNe. Indeed, SySts and PNe are well separated in two different loci. Only two objects are found to be misplaced in the region of young PNe (Hen~2--171 and H~1--36). In particular, H~1--36 is a well known D--type SySt with a variable Mira star \citep{Medina1995}. The misplacement of known SySts in the optical diagnostic diagram has previously been reported for other known SySts, such as Hen~2--147 and the extragalactic SySt (IC10 SySt-1); in addition of two new likely SySts found recently in the dwarf spheroidal galaxy NGC 205 \citep{Goncalves2014}. \cite{Gutierrez1995} have also mentioned that the separation of D--type SySts and young PNe is not always possible with this diagnostic diagram because of the high density nature of these two types of sources. The bifurcation of the data points in this diagram is the outcome of the different densities between the two groups of objects. For instance, SySts have a density several times higher than PNe and this results in higher attenuation of [O~{\sc iii}]~5007~\AA~emission to that of [O~{\sc iii}] 4363~\AA~emission. This is the reason for the emission and hence the clear separation of the SySts from the PNe in the optical diagnostic diagram.

%
\begin{figure*}[!]
\centering
\includegraphics[width=18.28cm \vspace{0.1cm}]{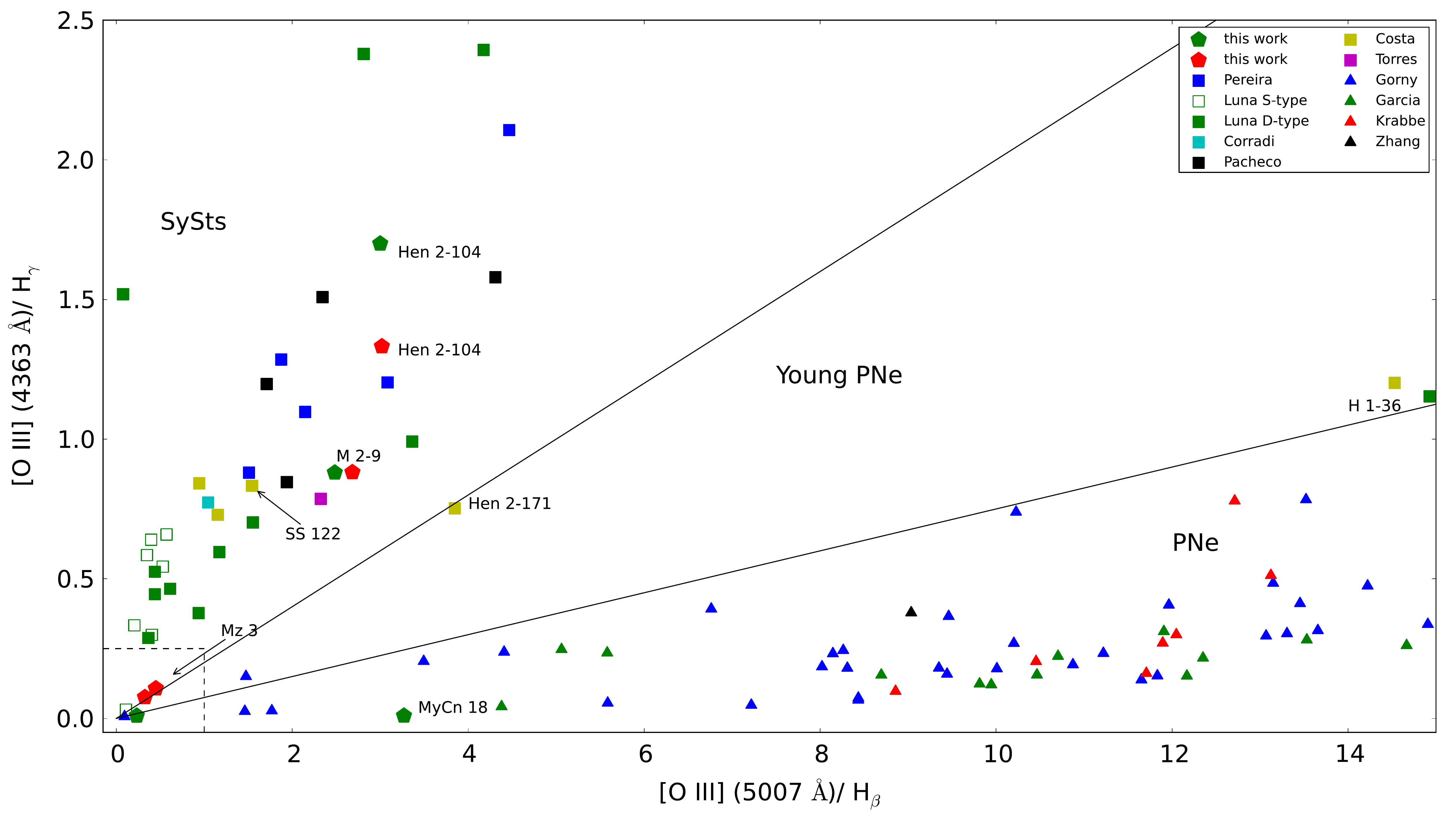} \vspace{0.5cm}
\includegraphics[width=18.4cm]{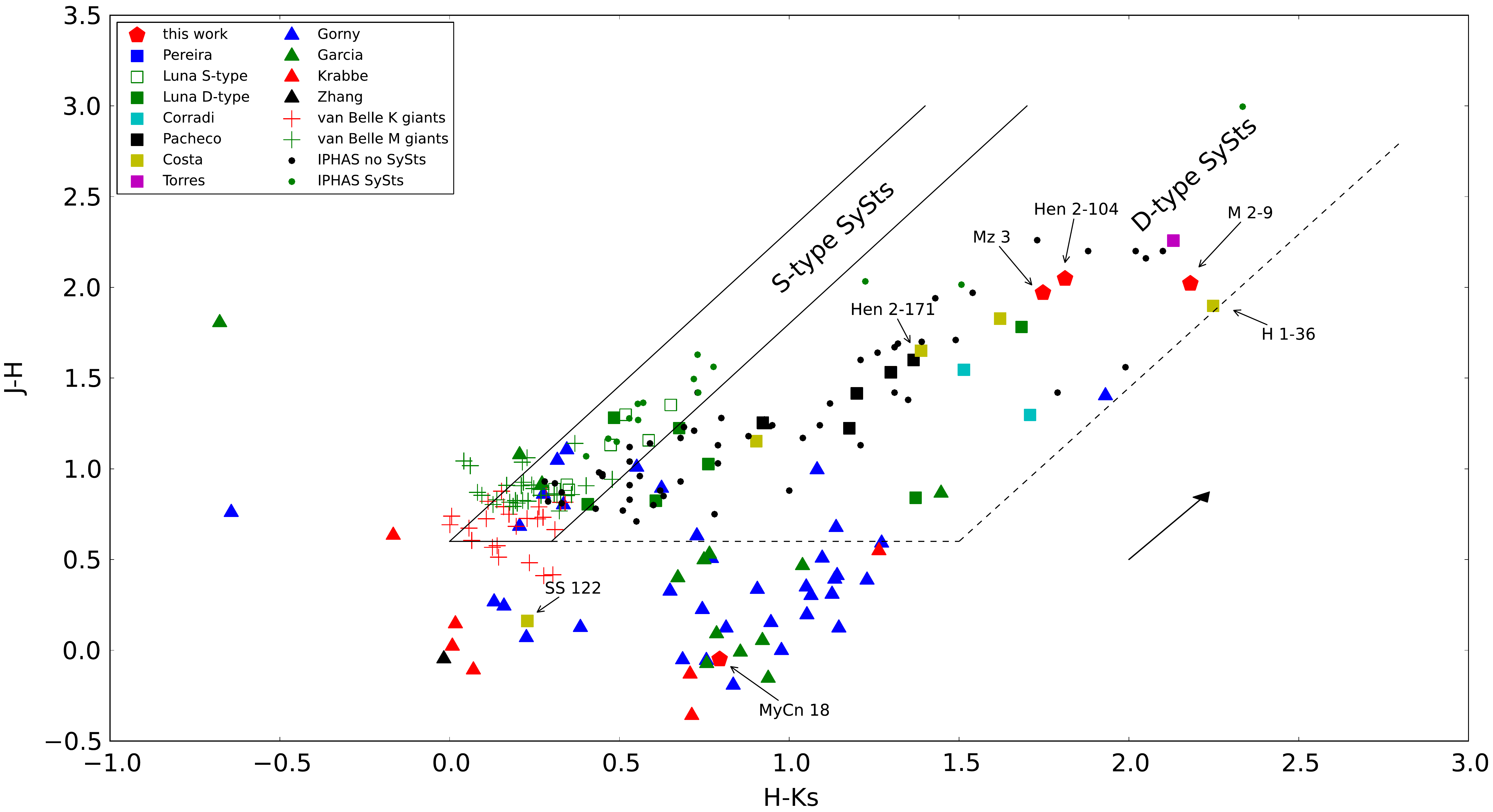}
\caption{$\lambda$4363/H$\gamma$ vs. $\lambda$5007/H$\beta$ (top) and 2MASS NIR colour-colour (bottom) diagnostic diagrams of PNe and SySts. The triangles and squares correspond to PNe and SySts, respectively. The red and green pentagons represent the bipolar nebulae, M~2--9, Mz~3, and Hen~2--104, with MyCn~18 as a special case. The colours represent the different sources (see Tables.~\ref{table:1} and \ref{table:2}) from which the data were obtained. The dashed box in the bottom left-hand corner of the optical diagnostic diagram defines a small region where the identification of the sources as SySts, Young PNe, or PNe is difficult due to the uncertainty in the weak [O~{\sc iii}] 4363~\AA~line, and the two diagonal lines correspond to an [O~{\sc iii}] 5007/4364 line ratio of 13.1 (upper line) and 27.4 (lower line). The red and green crosses in the 2MASS NIR diagram correspond to K and M giants respectively, and the black arrow corresponds to 3 mag extinction in the V band.}
\label{fig:fig1}
\end{figure*}
%

Regarding M~2--9 and Hen~2--104, these objects are well located in the regime of SySts, providing additional evidence for their symbiotic nature, whereas Mz~3 is placed in the bottom--left corner of the diagram, making its classification more uncertain. In particular, the emission-line ratios of Mz~3 have been estimated, separately, for its inner region and also for the outer bipolar lobes \citep{Smith2003}. The $\lambda$4363 \AA~line is found stronger in the inner region compared to the bipolar lobes, resulting in a higher $\lambda$4363/H$\gamma$, which moves Mz~3 deeper into the regime of SySts (see Fig.~\ref{fig:fig1}). However, its real nature still remains debatable. On the other hand, Hen~2--104 is a known D--type SySt with a Mira variable companion \citep{Whitelock1992}. As for M~2--9, several authors have classified this enigmatic object as a SySt but its true nature still remains debatable. Nevertheless, the presence of two compact circumstellar disks in M~2--9 have recently been confirmed \citep{Lykou2011,Castro2012}, providing further evidence of a likely binary system in its core, which is consistent with the SySt scenario, as much as with a genuine PNe with a binary central star.

Another PN once thought to be a SySt is MyCn~18 due to the similar characteristics it has with symbiotics. \cite{OConnor2000,Soker2012} and \cite{Clyne2014} suggested the possibility that MyCn~18 might have been a SySt, however, it is clearly placed in the PNe regime of the optical diagnostic diagram (Fig.~\ref{fig:fig1}). This further suggests that MyCn~18 is a genuine PN.

Also of importance are the uncertainties of the emission lines ([O~{\sc iii}] 4363~\AA, [O~{\sc iii}] 5007~\AA, H$\beta$, and H$\gamma$), which are lower than 15\% for most cases. However, for some PNe the uncertainty of the weak [O~{\sc iii}] 4363~\AA~line can be between 20--40\%. This makes objects located in the small region at the bottom left-hand corner (0 $<$ $\lambda$4363/H$\gamma$ $<$ 2.5 and 0 $<$ $\lambda$5007/H$\beta$ $<$ 1), and those bordering between regions, uncertain for proper identification. Mz~3 is located in the small boxed region, which makes identifying its symbiotic nature problematic. One additional cause that may affect the value of the 4363/H$\gamma$ line ratio, and thus the classification of the sources, is the Hg 4359~\AA~tulleric line, which is sometimes easily mistaken for the faint [O~{\sc iii}] 4363~\AA~line.   

%

\subsection{The 2MASS PNe versus SySts diagram}
\label{2mass}

Unlike the optical diagram just discussed, in the 2MASS \textit{J--H/H--Ks} diagnostic diagram the regime of SySts is mixed with other dusty celestial objects like Be, B[e], WR stars, as well as young dusty PNe and H {\sc ii} regions, making the identification of these objects difficult. Fig.~\ref{fig:fig1} also represents the 2MASS \textit{J--H/H--Ks} diagram for the same sample of SySts and PNe as in the optical plot just above it. This 2MASS diagram clearly shows that SySts define a specific region where all of them are placed. Only one SySt is found to be misplaced, namely \textit{SS~122}. This well known SySt is, however, found to be well placed in the regime of SySts in the optical diagram. We should also mention here that two SySts (Hen~2--171, H~1--36), which are found to lie in the region of young--PNe in the optical diagram, are also well placed in the regime of SySts in this infrared diagram.

According to their near-infrared colours and spectral energy distributions (SEDs), SySts are also classified into three subgroups; S-- (Stellar), D-- (Dusty), and D$^{\prime}$--type.
The S--type SySts show NIR temperatures in the range 2500--3000 K, which is consistent with the effective temperature of a K-- or M--type giant}, whereas D--type temperatures of 1000 K indicates a companion obscured within a warm dust shell \citep{Mikolajewska1997}. For comparison, we also plot the NIR colour values for a group of K and M giants \citep{Vanbelle1999}. These objects share the same regime with the S--type SySts, which is expected with a smaller range of J-H values. The difference between D-- and D$^{\prime}$--type SySts is the peak in their SED's, with the former and latter peaking at 5--15 $\upmu$m and 15--30 $\upmu$m respectively \citep{Ivison1995}. On the other hand, the group of PNe covers a larger range of \textit{H--Ks} colours from approximately $-0.7$ to 2.0, whilst the majority are concentrated in an even smaller region with \textit{J--H} colours between $-0.5$ and 0.5. There are, however, few PNe lying in the region of SySts.
  
M~2--9, Mz~3, and Hen~2--104, are shown as red pentagons in the infrared diagram, and lie within the region of D--type SySts. This is consistent with the Mira companion found in Hen~2--104 \citep{Whitelock1992}, but, thus far, there is no observational evidence of a Mira companion in either M~2--9 or Mz~3. The SEDs of these two objects indicates the presence of two stellar components, a hot and cold star \citep{Smith2003,Lykou2011}. In particular, the recently modelled SED of M~2--9, constructed by \cite{Lykou2011} using various photometric and spectroscopic data, shows a peak at $\sim$19 $\upmu$m, which is similar to those of D--type SySts. Nevertheless, this could simply prove the presence of a dense circumstellar envelope, not necessarily indicating that it is a SySt. Regarding the PN MyCn~18, it is found to be placed within the region where PNe are located in the 2MASS diagram, and if taken in conjunction with the optical diagram, only further implies its PN nature.

\subsection{H$\alpha$ broadening with double-peak profiles}

One particular velocity component that we did not reproduce in our models is the high, broadened velocity that emanates from the central region of each object (see the observed spectra in Figs.~\ref{fig:fig4}, \ref{fig:fig5}, and \ref{fig:henarrays}). However, we thought it would be of interest to analyse the velocity profiles that they exhibit. Interestingly, we found that the H$\alpha$ spectra from slit \textit{b} and \textit {c} of M~2--9, show a bright double-peaked profile, with the blue peak fainter than the red \citep{Lopez2012}. This kind of H$\alpha$ profile is commonly found in SySts, as are broad single peak profiles \citep{Winckel1993,Ivison1994}, and can be interpreted as the result of a strong (high-mass loss rate) and fast stellar wind expelled by the cold companion. The exact mechanism, however, remains undetermined.

%
\begin{figure}
\centering
\resizebox{\columnwidth}{!}{
{\includegraphics[width=0.5\columnwidth]{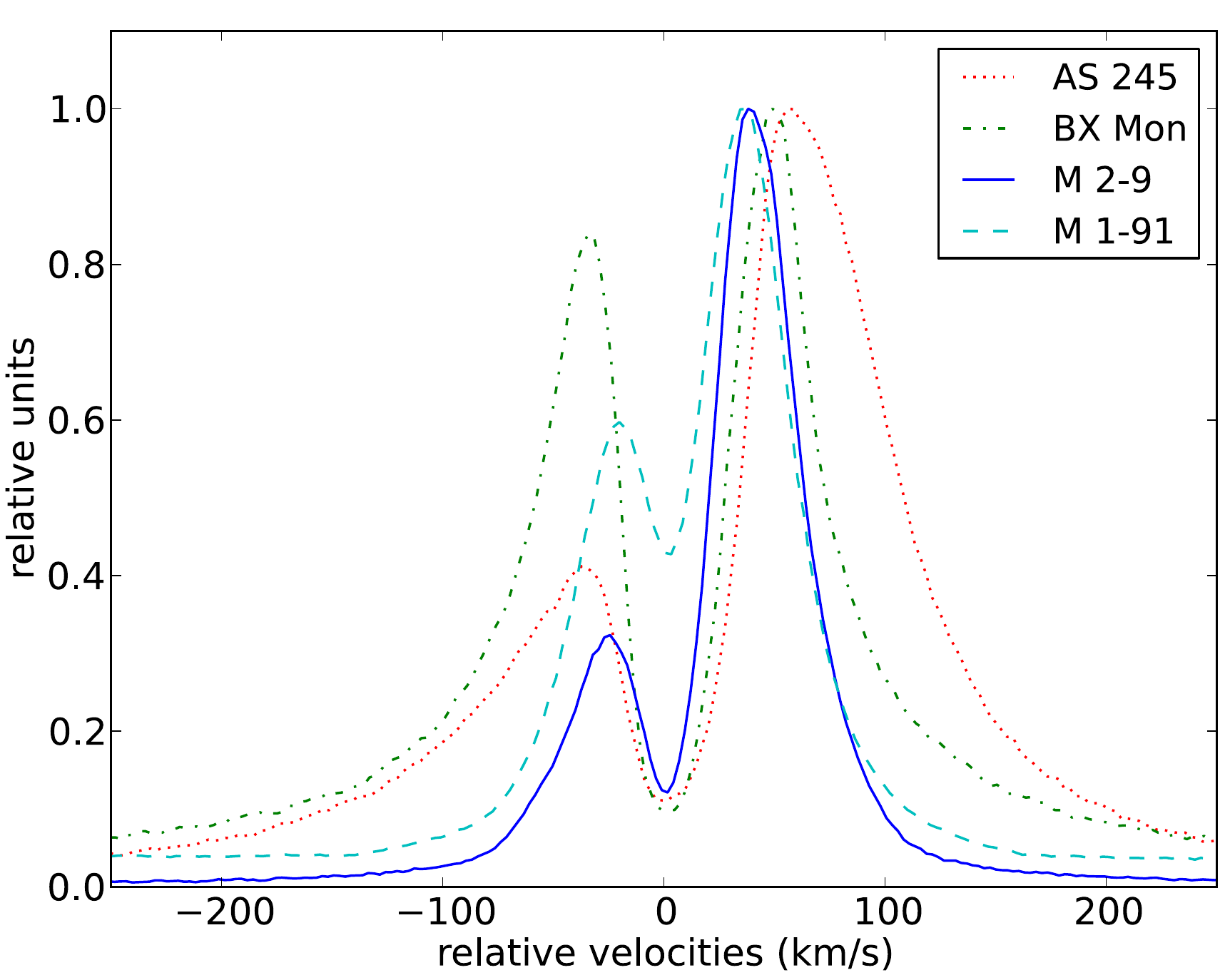}}}
\caption{H$\alpha$ line profiles of two known SySts, AS 245 and BX Mon, and two candidate SySts, M~2--9 and M~1--91. The data were obtained from the San Pedro M\'artir observatory (Akras et al. in prep). The x--axis represents the expansion velocities, where arbitrarily it has been given a zero velocity to the absorption dips for direct comparison.}
\label{fig:haprofiles}
\end{figure}
%

High-dispersion H$\alpha$ line profiles of two known SySts, BX~Mon and AS~245, and two candidate symbiotic nebulae, M~1--91 and M~2--9, are presented in Fig.~\ref{fig:haprofiles}. M~2--9 exhibits an absorption feature similar to what we find for the two known SySts, whereas it appears less absorbed in M~1--91, indicating a different degree of obscuration along our line of sight \citep[see][]{Torres-Peimbert2010}. Moreover, it may also be associated with the source type \citep[S-- or D--type;][]{Winckel1993} or possible H$\alpha$ variations \citep{Ivison1994}. The detection of the He {\sc ii} 6545 $\AA$ Raman scattered line in M~2--9 by \cite{Lee2001} in conjunction with the non-detection of any He recombination line suggests a highly obscured object. Moreover, M~2--9 and M~1--91 show significantly lower velocities (31 and 29 km s$^{-1}$) compared to those of BX~Mon and AS~245 (42 and 48 km s$^{-1}$), which is probably associated with the different evolutionary stages of the binary system and/or the presence of an extended Raman-scattering origin \citep{Lee2001}. It should also be mentioned here that the observed [O {\sc iii}] 5007/H$\beta$ and [O {\sc iii}] 4363/H$\gamma$ line ratios of M~1--91 \citep{Rodriguez2001} place its core in the regime of SySts in the optical diagnostic diagram, whereas the bipolar collimated lobes are placed in the region of young PNe, just like that of M~2--9.

In addition to this analysis, and using the SPM catalogue of galactic planetary nebulae \citep{Lopez2012}, we revised all the available high-dispersion echelle spectra for the known PNe that posses a binary system at their core. Two out of 26 PNe were found to exhibit H$\alpha$ double peak profiles with high velocities, namely Hen~2--171 and Sh~2--71. Surprisingly, Hen~2--171 is an already known SySt star, whereas Sh~2--71 is classified as a PN. Sh~2--71 can not be applied to the optical diagnostic diagram since there is no available emission line spectra from its core, however, its NIR colours place it firmly within the regime of SySts in the 2MASS diagram. Recently, \cite{Mocnik2015} confirm the binarity of Sh~2--71's central star, namely CS1, where the two companions are found to be a hot O--type subdwarf and a Be star.

\section{Discussion}
\label{discussion}   

The long-slit echelle spectra provided us with the necessary information about the 3-D structures of M~2--9, Mz~3, and Hen~2--104, allowing for the generation of self-consistent 3-D models.
The synthetic spectra generated from each model compares closely with their corresponding long-slit echelle spectra, which emphasises the accuracy of our models. 

\subsection{The morpho-kinematics of M~2--9.}

What makes M~2--9 intriguing is not only its detailed morphology but the peculiar behaviour within its internal region. The motion of the features N1 and S1 were initially thought to be knots moving in a lateral direction along the walls of the bulb, however, \cite{Corradi2011} described this apparent motion as excitation of the gas on the walls by a collimated spray of high velocity particles (jet) from the central source. In particular, \cite{Corradi2011} described that the walls of the inner cavity (bulb) are ionised by shocks produced by the impact of the jet particles. This would explain for the lighthouse effect seen in the bulb. \cite{Livio2001}, on the other hand, proposed a beam of ionising photons (light beam). A light beam would require most of the illuminating source to be obscured, and would represent an additional source of localised ionisation or scattered light. We accepted the former explanation by \cite{Corradi2011} as a more probable scenario for this unusual motion, which is why our model of M~2--9 includes surface brightness enhancements for N1 and S1 instead of physical components (like knots). Closely observing the shape of these features on the \textit{HST} [N~{\sc ii}] image, especially that of N1, it can be seen that they follow the curvature of the wall, which strongly suggests that they are the result of illumination by excitation. Modelling of these features (and those of N2, S2, N3, and S3) worked well using this approach, as shown in the synthetic spectra of Fig.~\ref{fig:fig4}.

It is also worth noting that the [N~{\sc ii}] emission from the central region is not reproduced in our model of M~2--9. This emission is the result of the gaseous interactions between the two binary stars at its core, with the hot companion ionising the stellar wind from the cold companion. This fast expanding ionised wind is responsible for the broad wings in the H$\alpha$ line, commonly found in SySts. The emission of M~2--9 has been found by \cite{Trammell1995} to be the combination of intrinsic/local emission and that of scattered emission from the central region. This scattered emission adds additional broadening to the emission lines for the N2 feature. This implies that we may slightly overestimate the real expansion velocities of the components. However, our {\sc shape} model of M~2--9 provides a very good interpretation of the current available data.

On another note, an energetic rotating jet \citep{Corradi2011} is most likely due to an existing binary system, which could explain for the tori or rings seen in \cite{Castro2012}. This is why our model includes equatorial ring-like structures. Both equatorial rings appear to fit the waists of their corresponding coaxial lobes (see {\sc shape} image of Fig.~\ref{fig:m29image}), which might explain for the formation of the bulb and neutral shell. \cite{Castro2012} also describes this system of rings as being ejected by one star, rather than a very late AGB or very early post-AGB phase, because of the small differences in their central offsets (0.34$^{\prime\prime}$) and systemic velocities (0.6 km s$^{-1}$). Therefore, the nebula would have been formed in, at least, two separate events. Also, the observed [O {\sc iii}] $\lambda$4363/H$\gamma$ and [O {\sc iii}] $\lambda$5007/H$\beta$ emission line ratios from its central region indicate the presence of a symbiotic binary system (see Fig.~\ref{fig:fig1}). Moreover, the double-peaked H$\alpha$ profile from its central region resembles those from other known SySts (Fig.~\ref{fig:haprofiles}), which provides further evidence of this scenario \citep{Torres-Peimbert2010}. See also \cite{Winckel1993} and \cite{Ivison1994} for a description of this.

Furthermore, the formation of the feathers is not so easy to explain. The extended emission appears to originate along the walls of the bulb, which suggests the possibility that it is a separate event that formed sometime after the bulb itself. Moreover, if the bulb formed after the feathers then its outflowing material could have interacted and moved along the walls of the feathers \citep{Kwok1982}, which would explain their similar appearance. 
A pair of high-velocity dust blobs (N4 and S4) can also be found far from the poles of the bulb along the nebula's axis of symmetry \citep[see Fig. 1 from][]{Corradi2011} and has extended emission reaching all the way out to them, therefore, the feathers and dust blobs possibly formed at the same time. We should also note that the dust blobs and the extended emission connecting them were not included in our model due to the unavailability of long-slit spectral data for these components. Finally, as mentioned in Sect.~\ref{m2-9}, our {\sc shape} fit for the outer lobes supports a scenario in which an initially radial expansion becomes redirected, presumably by dense equatorial material into a collimated, laterally expanding, outflow directed along the main symmetry axis of the nebula. This kind of recollimation is seen in outflows in star formation \citep{Rawlings2013}.

\subsection{The morpho-kinematics of Mz~3.}

Modelling the components of Mz~3 proved a challenge due to their highly complex morphologies. The bilobed shell BL1, its internal frothy textures (bubbles and sub-bubbles), and the protrusions seen at the tips of the shell were all carefully constructed, both morphology and kinematically, to match those seen in the observed high-quality long-slit spectra in Fig.~\ref{fig:fig5} and in the \textit{HST} image of Fig.~\ref{fig:mz3image}. To replicate these observed spectra from Gu04 a second pair of inner bilobes (bubble 1 and bubble 2) were included. This might suggest a relatively young ejection event due to the presence of a binary system at its centre. Interactions of this possible outflow with the apexes of BL1 might explain for the interrupted blistered regions and protrusions seen in the \textit{HST} image of Fig.~\ref{fig:mz3image}, however, we find that BL1 and these inner lobes share the kinematical age, which deviates from this hypothesis.   

Also of interest are the knots (initially detected by Gu04) found just above the poles of BL1. These are difficult to observe in the composite \textit{HST} and [N~{\sc ii}] images but the long-slit spectrum of Gu04 at P.A. 8$^{\circ}$ (where the slit goes through the symmetry axis and knots) of Fig.~\ref{fig:fig5} shows these knots to be present. In fact, the synthetic and observed spectra could not be correctly correlated without the inclusion of these components. Their kinematical ages, assuming particular parameters, indicates that they could have formed around the same time as BL2 and EE. Their origination, however, remains unclear.

BL3, on the other hand, was a difficult component to model because of the multitude, direction, and size of the stringy, ray-like features that emanate from the nebula's central region. Modelling these rays would not have been possible without the use of the detailed spectra from Gu04. What also makes BL3 interesting is the velocity by which these features expand. The rays were modelled with a Hubble-like expansion, which agrees very well with the observed spectra. We assumed a radial motion and an expanding velocity that is proportional to the distance from the nebular centre. As for the density distribution of each ray, we performed this visually using the observed spectra and \textit{HST} image in Fig.~\ref{fig:mz3image}.
Gu04 and SG04 described BL3 as probably the oldest part of the nebula because of its size, velocity, and its somewhat older kinematical age to those of BL1, BL2 and EE. Our analysis of BL3 further confirms their findings. 

We initially thought that the equatorial ellipse (EE), which is very faint in the composite \textit{HST} image, might be the earliest ejection event due to its large size and halo-like appearance, but we obtained a similar kinematical age to Gu04, which puts this component in the same age category as BL2 and the knots. This suggests that the knots, BL2, and EE could be related to each other due to the similarity of their ages.

\subsection{The morpho-kinematics of Hen~2--104.}

In contrast to the results of previous investigations by \cite{Corradi2001} and \cite{Santander2008}, we find that there is a considerable difference in the kinematical age of the two sets of lobes. This is manifest from the two different values for the constant of proportionality $k$ that determines the velocity field (see Table~\ref{table:henvels}). We find that the ratio between the velocities for the inner and outer pair of lobes is $6/2.5=2.4$, and is roughly consistent with the ratio of width between them. The ratio between the lengths, as given by the beginning and end of the region of knots and rings for the large and small lobes, respectively, is larger. We therefore conclude that the two sets of lobes are not coeval as proposed by the earlier studies, and suggest that the smaller set of lobes are roughly 2.4 times younger than the larger set.

The value of the constant $k=10$~km~s$^{-1}$~arcsec$^{-1}$ for the red knot that we identify with the jet corresponds to a kinematical age that is even younger than that of the smaller (inner) lobes. However, further study of the outer jets should be performed to test for their kinematical age.

The new detailed reconstruction of the distribution of knots, including the regions that are tangential to the line of sight, based on the long-slit spectra, suggests that the overall distribution of the knots of Hen~2--104 do not appear to be uniform over the azimuthal angle of the lobes. There are large regions with considerably fewer knots than in others. There is, however, no clear symmetry pattern between the two lobes regarding these variations in number density of knots and filaments. This result remains tentative, since the spectroscopic coverage in the tangential regions needs to be complete to make sure that no knots and filaments have been missed or misplaced. A better spectroscopic coverage with future observations in the tangential regions might help to strengthen or weaken this result.

From the modelling of the overall structure of the bipolar lobes, in the transition region from the smooth to filamentary emission distribution, we found a discontinuity in the opening-angle of the lobe. At the position where the knots and filaments start, the opening-angle increases suddenly by roughly 10$^{\circ}$ to 20$^{\circ}$ (see the mesh images for an example of this). This change in structure hints towards a change in thermal and velocity structure in the lobes, where a transition from a stable to an unstable shock structure happened. We speculate that the low-brightness region and brighter rings in the inner bipolar lobes might correspond to similar separate regions. In the future, the brighter rings might also break up into smaller scale knots and filaments due to thermal and/or dynamic instability.

Both on the red and blue side of the lobes we found a condensation that did not fit in the position-velocity profile established by the hourglass pattern. Each condensation is close to the symmetry axis of the nebula, and are about half-way from the centre to the edges of the inner hourglass lobes, approximately along the axis. It turns out that assuming they are located along the axis with a velocity of 105 km s$^{-1}$ produces
the correct placement in the synthetic spectra. Therefore, they might be residuals of the collimated jet ejecta. 

Finding the velocity of the outer jet features might also help to clarify whether they were ejected in a short burst or in a long duration process. In the case of a short burst, the space velocity of the outer jets should be proportional to their distance and about 3--4 times higher than the inner jet condensation. However, if a more continuous jet process produces these features, the velocities of the inner knot and outer jet features should be comparable.

\subsection{Separating symbiotic from planetary nebulae.}

The main reason for SySts and PNe being distinguished from each other in the optical diagram, as proposed by \cite{Gutierrez1995}, is their difference in physical conditions, especially the significantly lower core density in PNe. The same is true for all 35 recently discovered SySts in M~31 \citep[see Fig. 5 from][]{Mikolajewska2014}.

Due to their high densities, young PNe and SySts can occupy the same region in this diagram, as already pointed out by \cite{Gutierrez1995} and \cite{Pereira2005}. Also, the fact that at least two known SySts \citep[Galactic SySt Hen~2--147 and extragalactic IC10 SySt-1;][]{Mikolajewska1997b,Goncalves2008,Goncalves2015} are placed in the locus of the evolved PNe in the diagram, proves that there are pros and cons in the use of such diagnostic diagrams. So we should take the results indicated by them with caution. It is however worth noting that in the optical spectra of Hen~2--147 and IC10 SySt-1, as well as those of M~2--9 and Mz~3, the O {\sc vi} Raman scattered emission line at $\lambda\lambda$ 6830, 7088 were never detected. If so, no matter what the other properties of these nebulae are, they would be directly classified as SySts; since the presence of this emission line is a phenomenon uniquely found in SySts, not in all, but in at least 50\% of them \citep{Schmid1994}. Altogether these arguments seem to indicate that not only two (possibly and likely) SySts in NGC~205, but also other known SySts (Galactic and extragalactic) are misplaced in the optical diagram of Fig.~\ref{fig:fig1}, probably due to other (unknown) effects than the density.

Because of the obvious relevance of the NIR spectrum range to study the properties of the red-giant companion in a SySt, the way out of the above limitation on the use of purely optical spectrum is to use it simultaneously with the 2MASS diagram. Though a few PNe appear inside the locus of SySts in the 2MASS diagram of Fig.~\ref{fig:fig1}, only one SySt is outside the appropriate locus. Also, M~2--9 and Mz~3 are, once again, located well inside the locus of SySts, further suggesting their symbiotic nature.

As mentioned above, the presence of the O {\sc vi} Raman scattered emission line is a powerful criterion to classify a source as a SySt \citep[see, for instance,][]{Belczynski2000}. The broad emission features at around 6830 and 7088 $\AA$ are actually O {\sc vi} $\lambda\lambda$ 1032,1038 lines that are Raman-scattered by atomic hydrogen, where the scattering H atom initially in the 1s state de-excites to the 2s state after scattering. A He {\sc ii} Raman-scattered feature, around 4851 $\AA$ has been reported in the spectrum of the symbiotic nova RR Telescopii by \cite{VanGroningen1993}, and atomic physics tells us that it is also natural to expect that a stronger Raman-scattered He {\sc ii} line will be found around 6545 $\AA$, blueward of [N {\sc ii}] 6548~\AA, in the same object, and probably in other symbiotic stars \citep{Nussbaumer1989}. \cite{Lee2001} pointed out that the He {\sc ii} emission lines require much less H {\sc i} column densities for Raman scattering than the O {\sc vi}, and that we may expect that the former Raman scattering may be observed in broader classes of objects than symbiotic stars. 

In the particular case of M~2--9, the He {\sc ii} Raman scattering around 6545 $\AA$ was detected, and the broad H$\alpha$ wings of this nebula was studied and attributed, as much as in known SySts (RR Tel and He~2--106) to the hypothesis of having a Raman-scattering origin \citep{Lee2001}. As the latter authors discuss, at least one young PNe also presents the He {\sc ii} Raman scattering around 4851 $\AA$, NGC 7027 \citep{Pequignot1997}.

Furthermore, the broad H$\alpha$ lines with a double or even single peak profile \citep{Winckel1993} may
be useful in understanding the physical conditions at the core of M~2--9, Mz~3, Hen~2--104, and even M~1--91. Hen~2--104 shows a broad single profile similar to those of known D--type SySts such as RR~Tel, Rx~Pup, and Hen~2--171 \citep{Winckel1993}, whereas M~2--9 and M~1--91 show double-peaked H$\alpha$ profiles similar to S--type SySts.

Due to the difference in evolutionary stages between PNe and SySts, different morphological characteristics may be expected, such as in the jets that are observed in both type of system. A study of their kinematical ages show that they, more often than not, predate the kinematical ages of bipolar lobes associated with PNe \citep{Jones2014}, such that we would expect younger jets to be observable in SySts. This hypothesis is supported in this work by the detection of a jet feature close to the binary centre of Hen~2--104. For these binary cores in SySts it is not unusual for long periods $\geq$ 10 years. For SySts and PNe with binary cores, depending on mass ratios and rates of stellar evolution of individual systems, will evolve into nova systems or the appropriate analogous sub-type of cataclysmic variable for their characteristics. However, further investigation is still necessary to prove these hypotheses and provide a robust technique  for separating SySts from young PNe.

\section{Conclusions}
\label{conclusions}

We have carried out a morpho-kinematic and spectroscopic study of M~2--9, Mz~3, and Hen~2--104 in order to further understand their structures, kinematics, and nature. The morphology and kinematics of these objects have been accurately constrained by the use of detailed long-slit spectra. The velocities of the individual structures for each object have been determined, and a velocity field was fitted to these to reveal the red- and blue-shifted motion of their material. Kinematical ages have also been obtained for the separate components of M~2--9 and Mz~3.

The laterally rotating features of M~2--9 were modelled as surface brightness enhancements, which are most likely the result of excitation on the walls of the bulb component by a collimated spray of high velocity particles. On the other hand, the morpho-kinematic modelling of Mz~3 has revealed a second inner bipolar structure within BL1 and a set of knots found just above the apexes of BL1. For Hen~2--104, we found what appears to be two knotty jet structures close to the symmetry axis and found at $\sim$11$^{\prime\prime}$ and 13$^{\prime\prime}$ on either side of the outer bipolar lobes. Also, according to our modelling of Hen~2--104, the kinematical age of the inner and outer bipolar lobes are different, with the outer lobes 2.4 times older.

Apart from just studying the morphological and kinematical characteristics, {\sc shape} also provided us with the density distribution of each nebula in the form of an ASCII data cube, which will allow for the future work of creating 3-D photo-ionisation models. These data cubes can be used as an input for other external codes such as {\sc moccasin} \citep{Ercolano2003} to create complex 3-D photo-ionisation models and investigate, in detail, the chemical structure of the nebulae, while avoiding the time consuming task of reproducing the density distribution of these models from the beginning.

The optical emission line ratios and 2MASS colour values present further evidence for the symbiotic nature of M~2--9 and Mz~3. Both diagnostic diagrams have been shown to successfully separate a multitude of known SySts from genuine PNe. However, the NIR colour-colour diagram can not be used for the classification of SySts since several other objects such as Be, B[e], WR, young PNe, and H {\sc ii} regions, share the same range of colours.

The H$\alpha$ profile of M~2--9 shows a broad line with an absorption dip and two asymmetric components that are usually found in SySts because of the fast stellar wind and the high mass-loss rate of the cold companion. This is indicative of a binary symbiotic system. 

%
 
\begin{acknowledgements}
We gratefully acknowledge the support of the College of Science, NUI Galway (under their PhD fellowship scheme).
Based on observations made with the NASA/ESA Hubble Space Telescope, obtained from the data archive at the Space Telescope Institute. This paper makes use of data obtained from the San Pedro M\'artir kinematic catalogue of Galactic Planetary Nebulae. S.A. is supported by CAPES post-doctoral fellowship ``Young Talents Attraction'' - Science Without Borders, A035/2013. W.S. acknowledges funding from grant UNAM-DGAPA PAPIIT IN101014. E.H. is supported by funding from the Irish Research Council under their postgraduate scheme. We thank the anonymous referee for their thorough review and highly appreciate the comments and suggestions, which significantly contributed to improving the quality of the publication.
\end{acknowledgements}

%
%

\bibliographystyle{aa}
\bibliography{reference_database}

\pagebreak{}
\section*{Appendix A: Tables and Mesh Images}

The optical emission line intensities for the various SySts and PNe are shown in Tables.~\ref{table:optss} and \ref{table:optpne} respectively. The 2MASS NIR colour magnitude values for the same sample of SySts and PNe are presented in Tables.~\ref{table:2massnir} and \ref{table:2masspne}. Shown in Fig.~\ref{fig:henvelocityfield} are the 3-D {\sc shape} images of M~2--9, Mz~3, and Hen~2--104 respectively. In all of these images, the red line represents the top view, the blue line represents the right (or side) view, and the green line represents the symmetry (or major) axis.

%
\begin{table*}
\caption{\label{table:optss}Line intensities for the nucleus of specific symbiotic nebulae from various literature}             
\label{table:1}      
\centering          
\begin{tabular}{c c c c c c c} 
\hline\hline\addlinespace[3pt]       
                       
Object & Type & {[H{\sc$\gamma$}]4340~\AA} & {[O~{\sc iii}]4363~\AA} & {[H{\sc$\beta$}]4861~\AA} & {[O~{\sc iii}]5007~\AA} & Reference \\ \addlinespace[1pt]
\hline\addlinespace[2.5pt]
   AS210 & SySt & 64.3 & 77.0 & 100 & 171.1 & \cite{Pacheco1992} \\
   Hen 2-38 & SySt & 53.9 & 220.1 & 100 & 707.7 & \cite{Pacheco1992} \\
   RR Tel & SySt & 50.6 & 42.8 & 100 & 193.8 & \cite{Pacheco1992} \\
   RX Pup & SySt & 54.5 & 82.2 & 100 & 234.4 & \cite{Pacheco1992} \\
   HM Sge & SySt & 45.2 & 71.4 & 100 & 430.9 & \cite{Pacheco1992} \\
   He 2-106 & SySt & 51.2 & 43.1 & 100 & 94.3 & \cite{Costa1994} \\
   He 2-127 & SySt & 45.0 & 32.8 & 100 & 115.4 & \cite{Costa1994} \\
   He 2-171* & SySt & 34.3 & 25.8 & 100 & 384.8 & \cite{Costa1994} \\
   H 1-36* & SySt & 46.1 & 55.4 & 100 & 1453.0 & \cite{Costa1994} \\
   SS 122 & SySt & 56.9 & 47.4 & 100 & 154.3 & \cite{Costa1994} \\
   He 2-25 & SySt & 44.4 & 161.6 & 100 & 475.9 & \cite{Corradi1995} \\
   Th 2-B & SySt & 36.2 & 28.0 & 100 & 104.4 & \cite{Corradi1995} \\                   
   He 2-171 & SySt & 45.1 & 95.0 & 100 & 446.6 & \cite{Pereira1995} \\  
   He 2-127 & SySt & 46.3 & 50.8 & 100 & 214.7 & \cite{Pereira1995} \\
   SS 122 & SySt & 35.0 & 30.8 & 100 & 150.9 & \cite{Pereira1995} \\
   He 2-106 & SySt & 44.8 & 53.9 & 100 & 308.5 & \cite{Pereira1995} \\
   Hen 2-104$^{\dagger}$ & SySt & 39.7 & 51.0 & 100 & 187.5 & \cite{Pereira1995} \\
   Hen 2-104$^{\dagger}$ & SySt & 53.9 & 71.8 & 100 & 301.9 & \cite{Pacheco1996} \\
   He 2-171 & SySt & 45.0 & 44.6 & 100 & 336.4 & \cite{Luna2005} \\
   SS 122 & SySt & 38.6 & 23.0 & 100 & 117.2 & \cite{Luna2005} \\
   H 2-38 & SySt & 47.3 & 113.2 & 100 & 417.5 & \cite{Luna2005} \\
   H 2-5 & SySt & 41.2 & 24.1 & 100 & 34.9 & \cite{Luna2005} \\
   H 1-36** & SySt & 45.1 & 52.0 & 100 & 1492.7 & \cite{Luna2005} \\
   H 1-25 & SySt & 40.2 & 21.1 & 100 & 43.9 & \cite{Luna2005} \\
   Hen 3-1342* & SySt & 44.6 & 1.4 & 100 & 11.0 & \cite{Luna2005} \\
   H 2-34 & SySt & 36.8 & 20.0 & 100 & 52.8 & \cite{Luna2005} \\
   Hen 3-1591 & SySt & 49.9 & 118.7 & 100 & 281.2 & \cite{Luna2005} \\
   Hen 3-1761 & SySt & 54.2 & 35.7 & 100 & 57.1 & \cite{Luna2005} \\
   Hen 3-1410 & SySt & 43.1 & 12.9 & 100 & 40.5 & \cite{Luna2005} \\
   Th 3-29 & SySt & 45.0 & 28.8 & 100 & 39.6 & \cite{Luna2005} \\
   Wray 16-377 & SySt & 44.1 & 14.7 & 100 & 20.6 & \cite{Luna2005} \\
   SS 96 & SySt & 38.9 & 17.3 & 100 & 43.9 & \cite{Luna2005} \\
   SS 71 & SySt & 41.3 & 11.9 & 100 & 36.6 & \cite{Luna2005} \\
   AS210 & SySt & 44.5 & 16.8 & 100 & 93.6 & \cite{Luna2005} \\
   AS269 & SySt & 37.2 & 56.5 & 100 & 7.7 & \cite{Luna2005} \\
   RR Tel & SySt & 48.3 & 33.9 & 100 & 155.2 & \cite{Luna2005} \\
   K 66 & SySt & 30.8 & 14.3 & 100 & 61.2 & \cite{Luna2005} \\
\hline                  
\end{tabular}
\tablefoot{A list of symbiotics chosen from literature with extinction correction intensity. $^{\dagger}$objects of interest within this paper, *objects bordering between regions in the optical diagnostic diagram, **objects that fall outside their associated (SySt) region.}
\end{table*}
%

%
\begin{table*}
\caption{\label{table:optpne}Line intensities for the nucleus of specific PNe from various literature}             
\label{table:2}      
\centering          
\begin{tabular}{c c c c c c c} 
\hline\hline\addlinespace[3pt]       
                       
Object & Type & {[H{\sc$\gamma$}]4340~\AA} & {[O~{\sc iii}]4363~\AA} & {[H{\sc$\beta$}]4861~\AA} & {[O~{\sc iii}]5007~\AA} & Reference \\ \addlinespace[1pt]
\hline\addlinespace[2.5pt]
   MyCn 18 & PN & 46.8 & 0.45 & 100 & 327$^{a}$ & \cite{Kingsburgh1994} \\
   Mz 3$^{\dagger}$* & PN & 52.5 & 3.984 & 100 & 32.58 & \cite{Zhang2002} \\
   Mz 3$^{\dagger}$ & PN & 46.2 & 4.963 & 100 & 44.96 & \cite{Smith2003} \\                         
   NGC 1535 & PN & 46.8 & 12.65 & 100 & 1189.38 & \cite{Krabbe2006} \\
   NGC 2438 & PN & 46.8 & 9.56 & 100 & 1045.42 & \cite{Krabbe2006} \\
   NGC 2440 & PN & 46.8 & 23.98 & 100 & 1312.31 & \cite{Krabbe2006} \\
   NGC 3132 & PN & 46.8 & 4.56 & 100 & 885.72 & \cite{Krabbe2006} \\
   NGC 3242 & PN & 46.8 & 14.05 & 100 & 1204.84 & \cite{Krabbe2006} \\
   NGC 6302 & PN & 46.8 & 36.46 & 100 & 1271.16 & \cite{Krabbe2006} \\
   NGC 7009 & PN & 46.8 & 7.57 & 100 & 1170.75 & \cite{Krabbe2006} \\
   NGC 7027 & PN & 47.6 & 25.4 & 100 & 1657.9 & \cite{Zhang2007} \\
   M 2-9$^{\dagger}$ & PN & 53.1 & 46.8 & 100 & 268.4 & \cite{Torres-Peimbert2010} \\
   M 1-91 & PN & 45.9 & 36.1 & 100 & 232.5 & \cite{Torres-Peimbert2010} \\
   NGC 2392 & PN & 39.35 & 14.9 & 100 & 903.5 & \cite{Zhang2012} \\
   PB 5 & PN & 47.17 & 22.39 & 100 & 1422.24 & \cite{Gorny2014} \\
   PB 3 & PN & 48.24 & 14.26 & 100 & 1306.84 & \cite{Gorny2014} \\
   He 2-36 & PN & 47.26 & 19.46 & 100 & 1345.42 & \cite{Gorny2014} \\
   He 2-39 & PN & 47.24 & 22.88 & 100 & 1314.91 & \cite{Gorny2014} \\
   He 2-63 & PN & 47.09 & 12.67 & 100 & 1020.17 & \cite{Gorny2014} \\
   He 2-67 & PN & 47.28 & 6.55 & 100 & 1165.17 & \cite{Gorny2014} \\
   He 2-70 & PN & 47.15 & 34.84 & 100 & 1022.69 & \cite{Gorny2014} \\
   He 2-84 & PN & 47.95 & 11.19 & 100 & 1121.92 & \cite{Gorny2014} \\
   He 2-85 & PN & 47.03 & 14.81 & 100 & 1365.70 & \cite{Gorny2014} \\
   He 2-86 & PN & 46.64 & 3.42 & 100 & 843.42 & \cite{Gorny2014} \\
   He 2-96 & PN & 46.60 & 3.08 & 100 & 843.37 & \cite{Gorny2014} \\
   My 60 & PN & 47.90 & 19.46 & 100 & 1196.34 & \cite{Gorny2014} \\
   Pe 1-1 & PN & 48.26 & 7.67 & 100 & 944.27 & \cite{Gorny2014} \\
   ESO 095-1 & PN & 47.13 & 20.88 & 100 & 1538.56 & \cite{Gorny2014} \\
   ESO 320-2 & PN & 47.46 & 9.71 & 100 & 349.44 & \cite{Gorny2014} \\
   BMPJ1128-6 & PN & 47.32 & 37.08 & 100 & 1352.23 & \cite{Gorny2014} \\
   NGC 3918 & PN & 46.65 & 20.98 & 100 & 1578.26 & \cite{Gorny2014} \\
   K 1-23 & PN & 46.96 & 8.48 & 100 & 934.96 & \cite{Gorny2014} \\
   Wray 16-128 & PN & 47.63 & 11.32 & 100 & 440.80 & \cite{Gorny2014} \\
   MeWe 1-3 & PN & 47.46 & 18.61 & 100 & 676.26 & \cite{Gorny2014} \\
   MaC 1-2 & PN & 47.03 & 15.85 & 100 & 1490.70 & \cite{Gorny2014} \\
   He 2-102 & PN & 46.74 & 8.33 & 100 & 1000.89 & \cite{Gorny2014} \\
   He 2-103 & PN & 52.47 & 9.72 & 100 & 802.13 & \cite{Gorny2014} \\
   He 2-109 & PN & 46.93 & 9.04 & 100 & 1087.32 & \cite{Gorny2014} \\
   He 2-115 & PN & 46.68 & 2.58 & 100 & 558.55 & \cite{Gorny2014} \\
   He 2-117 & PN & 46.54 & 2.22 & 100 & 721.78 & \cite{Gorny2014} \\
   He 2-133 & PN & 52.81 & 8.08 & 100 & 1183.33 & \cite{Gorny2014} \\
   NGC 5939 & PN & 47.27 & 17.28 & 100 & 946.07 & \cite{Gorny2014} \\
   NGC 6326 & PN & 46.52 & 14.14 & 100 & 1330.61 & \cite{Gorny2014} \\
   StWr 4-10 & PN & 47.13 & 10.95 & 100 & 814.49 & \cite{Gorny2014} \\
   K 3-31 & PN & 50.25 & 13.13 & 100 & 1466.56 & \cite{Garcia2014} \\
   K 3-56 & PN & 47.38 & 11.71 & 100 & 506.08 & \cite{Garcia2014} \\
   K 3-87 & PN & 47.30 & 11.11 & 100 & 558.00 & \cite{Garcia2014} \\
   K 4-41 & PN & 46.65 & 5.63 & 100 & 994.52 & \cite{Garcia2014} \\
   M 2-50 & PN & 47.09 & 14.67 & 100 & 1190.65 & \cite{Garcia2014} \\
   Bl 2-1 & PN & 52.03 & 6.44 & 100 & 981.25 & \cite{Garcia2014} \\
   Hen 2-440 & PN & 46.08 & 1.96 & 100 & 437.94 & \cite{Garcia2014} \\
   NGC 6807 & PN & 46.86 & 13.19 & 100 & 1353.14 & \cite{Garcia2014} \\
   M 1-69 & PN & 46.78 & 7.12 & 100 & 1216.84 & \cite{Garcia2014} \\
   IC 4846 & PN & 46.84 & 7.31 & 100 & 1046.40 & \cite{Garcia2014} \\
   K 4-8 & PN & 46.90 & 10.15 & 100 & 1235.16 & \cite{Garcia2014} \\
   Sa 2-237 & PN & 46.99 & 10.47 & 100 & 1070.36 & \cite{Garcia2014} \\
   MaC 1-11 & PN & 49.07 & 7.64 & 100 & 869.54 & \cite{Garcia2014} \\
\hline                  
\end{tabular}
\tablefoot{A list of planetary nebulae (PN) chosen from literature with extinction correction intensity. $^{\dagger}$objects of interest within this paper, *objects bordering between regions in the optical diagnostic diagram. $^{a}$\,the intensity of the {[O~{\sc iii}]5007~\AA} line is obtained from the {[O~{\sc iii}]4959~\AA} line multiplied by 3.}
\end{table*}
%

%
\begin{table*}
\caption{\label{table:2massnir} 2MASS NIR (\textit{HJKs}) colour magnitude values of various SySts.}             
\label{table:3}      
\centering          
\begin{tabular}{c l c l c l c l c} 
\hline\hline\addlinespace[3pt]       
                       
Object & & Type & & {\textit{H}} & & {\textit{J}} & & {\textit{Ks}} \\ \addlinespace[1pt]
\hline\addlinespace[2.5pt]                   
   He 2-171 & & SySt & & 9.679 & & 8.028 & & 6.640  \\
   He 2-127 & & SySt & & 10.585 & & 9.433 & & 8.530 \\
   SS 122** & & SySt & & 9.958 & & 9.796 & & 9.567  \\
   Hen 2-104$^{\dagger}$ & & SySt & & 11.005 & & 8.957 & & 7.145 \\
   He 2-106 & & SySt & & 9.584 & & 7.756 & & 6.135  \\
   He 2-171 & & SySt & & 9.679 & & 8.028 & & 6.640 \\
   H 2-38 & & SySt & & 8.266 & & 7.240 & & 6.478  \\
   H 2-5 & & SySt & & 7.161 & & 6.030 & & 5.556  \\
   H 1-36* & & SySt & & 11.771$^{a}$ & & 9.873 & & 7.625  \\
   H 1-25 & & SySt & & 10.415 & & 8.633 & & 6.949 \\
   Hen 3-1342 & & SySt & & 9.605 & & 8.693 & & 8.348 \\
   H 2-34 & & SySt & & 9.423 & & 8.125 & & 7.607 \\
   Hen 3-1591 & & SySt & & 10.504 & & 9.680 & & 9.073 \\
   Hen 3-1761 & & SySt & & 6.681 & & 5.797 & & 5.446 \\
   Hen 3-1410 & & SySt & & 10.315 & & 9.158 & & 8.572 \\
   Th 3-29 & & SySt & & 8.924 & & 7.572 & & 6.921 \\
   Wray 16-377 & & SySt & & 9.802 & & 8.914 & & 8.605 \\
   SS73 96 & & SySt & & 8.187 & & 6.905 & & 6.421 \\
   SS73 71 & & SySt & & 11.374 & & 10.569 & & 10.162 \\
   AS210 & & SySt & & 9.087 & & 7.555 & & 6.256 \\
   AS269 & & SySt & & 12.148 & & 11.308 & & 9.936 \\
   RR Tel & & SySt & & 7.302 & & 6.079 & & 4.902 \\
   K 6-6 & & SySt & & 9.423 & & 8.199 & & 7.523 \\
   He 2-25 & & SySt & & 13.412 & & 12.115 & & 10.406 \\
   Th 2-B & & SySt & & 12.951 & & 11.405 & & 9.891 \\
   RX Pup & & SySt & & 6.397 & & 4.981 & & 3.782 \\
   HM Sge & & SySt & & 7.620 & & 6.020 & & 4.654 \\
\hline                  
\end{tabular}
\tablefoot{The data were obtained from the 2MASS All-Sky Catalog of Point Sources \citep{Cutri2003}. $^{a}$\,uncertain value. $^{\dagger}$objects of interest within this paper, *objects bordering between regions in the 2MASS colour-colour diagram, **objects that fall outside their associated (SySt) region.}
\end{table*}
%

%
\begin{table*}
\caption{\label{table:2masspne} 2MASS NIR (\textit{HJKs}) colors of various PNe.}             
\label{table:4}      
\centering          
\begin{tabular}{c l c l c l c l c} 
\hline\hline\addlinespace[3pt]       
                       
Object & & Type & & {\textit{H}} & & {\textit{J}} & & {\textit{Ks}} \\ \addlinespace[1pt]
\hline\addlinespace[2.5pt]                         
   NGC 1535 & & PN & & 12.539 & & 12.647 & & 12.577 \\
   NGC 2438 & & PN & & 12.672 & & 12.526 & & 12.509 \\
   NGC 2440 & & PN & & 15.671 & & 15.037 & & 15.203 \\
   NGC 3132 & & PN & & 9.754 & & 9.731 & & 9.723 \\
   NGC 3242 & & PN & & 12.102 & & 12.461 & & 11.748 \\
   NGC 6302 & & PN & & 11.255 & & 11.706 & & 9.442 \\
   NGC 7009$^{\dagger}$ & & PN & & 9.325 & & 9.456 & & 8.748  \\
   NGC 7027 & & PN & & 9.613 & & 8.37 & & 7.443 \\
   NGC 2392 & & PN & & 10.872 & & 10.919 & & 10.936  \\
   MyCn 18 & & PN & & 11.554 & & 11.603 & & 10.808 \\
   Mz 3$^{\dagger}$ & & PN & & 9.352 & & 7.355 & & 5.608 \\
   PB 5 & & PN & & 12.488 & & 11.897 & & 10.625 \\
   PB 3 & & PN & & 13.899 & & 13.703 & & 12.651 \\
   He 2-36 & & PN & & 9.867 & & 9.623 & & 9.463 \\
   He 2-39 & & PN & & 13.474 & & 12.614 & & 12.338 \\
   He 2-63 & & PN & & 15.55 & & 15.427 & & 14.613 \\
   He 2-67 & & PN & & 13.43 & & 13.431 & & 12.454 \\
   He 2-70 & & PN & & 12.544 & & 11.497 & & 11.180 \\
   He 2-84 & & PN & & 14.527 & & 13.422 & & 13.077 \\
   He 2-85 & & PN & & 15.198 & & 14.305 & & 13.681 \\
   He 2-86 & & PN & & 12.155 & & 11.853 & & 10.789 \\
   He 2-96 & & PN & & 12.577 & & 12.228 & & 11.178 \\
   My 60$^{\dagger\dagger}$ & & PN & & 12.604 & & 12.379 & & 11.635 \\
   Pe 1-1 & & PN & & 12.457 & & 12.065 & & 10.931 \\
   ESO 095-1 & & PN & & 15.201 & & 13.798 & & 11.867 \\
   ESO 320-2 & & PN & & 15.166 & & 14.157 & & 13.606 \\
   BMP J1128-6121 & & PN & & 14.155 & & 13.478 & & 12.340 \\
   NGC 3918 & & PN & & 11.691 & & 11.568 & & 10.422 \\
   K 1-23 & & PN & & 16.701 & & 15.941 & & 16.584 \\
   Wray 16-128 & & PN & & 15.761 & & 15.078 & & 14.872 \\
   MaC 1-2 & & PN & & 13.713 & & 13.446 & & 13.315 \\
   He 2-102 & & PN & & 14.636 & & 14.31 & & 13.661 \\
   He 2-103 & & PN & & 14.971 & & 14.34 & & 13.612 \\
   He 2-108 & & PN & & 12.173 & & 12.103 & & 11.877 \\
   He 2-109 & & PN & & 16.338 & & 15.832 & & 15.061$^{a}$ \\
   He 2-115 & & PN & & 12.033 & & 11.724 & & 10.598 \\
   He 2-117 & & PN & & 11.811$^{a}$ & & 11.424$^{a}$ & & 10.195$^{a}$ \\
   He 2-133 & & PN & & 12.132 & & 11.72 & & 10.579 \\
   NGC 5939 & & PN & & 14.777 & & 13.782 & & 12.700$^{a}$ \\
   NGC 6326 & & PN & & 14.747 & & 14.939 & & 14.104 \\
   StWr 4-10 & & PN & & 14.335 & & 14.209 & & 13.824  \\
   M 2-9$^{\dagger}$ & & PN & & 11.198 & & 9.177 & & 6.996 \\
   M 1-91 & & PN & & 14.099 & & 11.841 & & 9.710 \\
   K 3-31 & & PN & & 13.806 & & 13.339 & & 12.300 \\
   K 3-56 & & PN & & 15.973 & & 14.895$^{a}$ & & 14.689 \\
   K 3-87 & & PN & & 16.214 & & 15.684 & & 14.919  \\
   K 4-41 & & PN & & 15.602$^{a}$ & & 13.796$^{a}$ & & 14.473 \\
   M 2-50 & & PN & & 15.333 & & 15.242 & & 14.456 \\
   Bl 2-1 & & PN & & 13.473 & & 12.607 & & 11.16 \\
   Hen 2-440 & & PN & & 13.603 & & 12.686 & & 12.414 \\
   NGC 6807 & & PN & & 12.753 & & 12.826 & & 12.069 \\
   M 1-69 & & PN & & 13.261 & & 13.207 & & 12.286 \\
   IC 4846 & & PN & & 12.681 & & 12.835 & & 11.897 \\
   K 4-8 & & PN & & 14.792 & & 14.802 & & 13.946 \\
   Sa 2-237 & & PN & & 13.065 & & 12.566 & & 11.817 \\
   MaC 1-11 & & PN & & 14.179 & & 13.779 & & 13.107 \\
\hline                  
\end{tabular}
\tablefoot{The data have been obtained from the 2MASS All-Sky Catalog of Point Sources \citep{Cutri2003}. $^{\dagger}$objects of interest within this paper. $^{\dagger\dagger}$from \cite{Skrutskie2006}. $^{a}$\,uncertain value.}
\end{table*}
%

%
\begin{figure*}[ht]
\centering
\includegraphics[width=18cm]{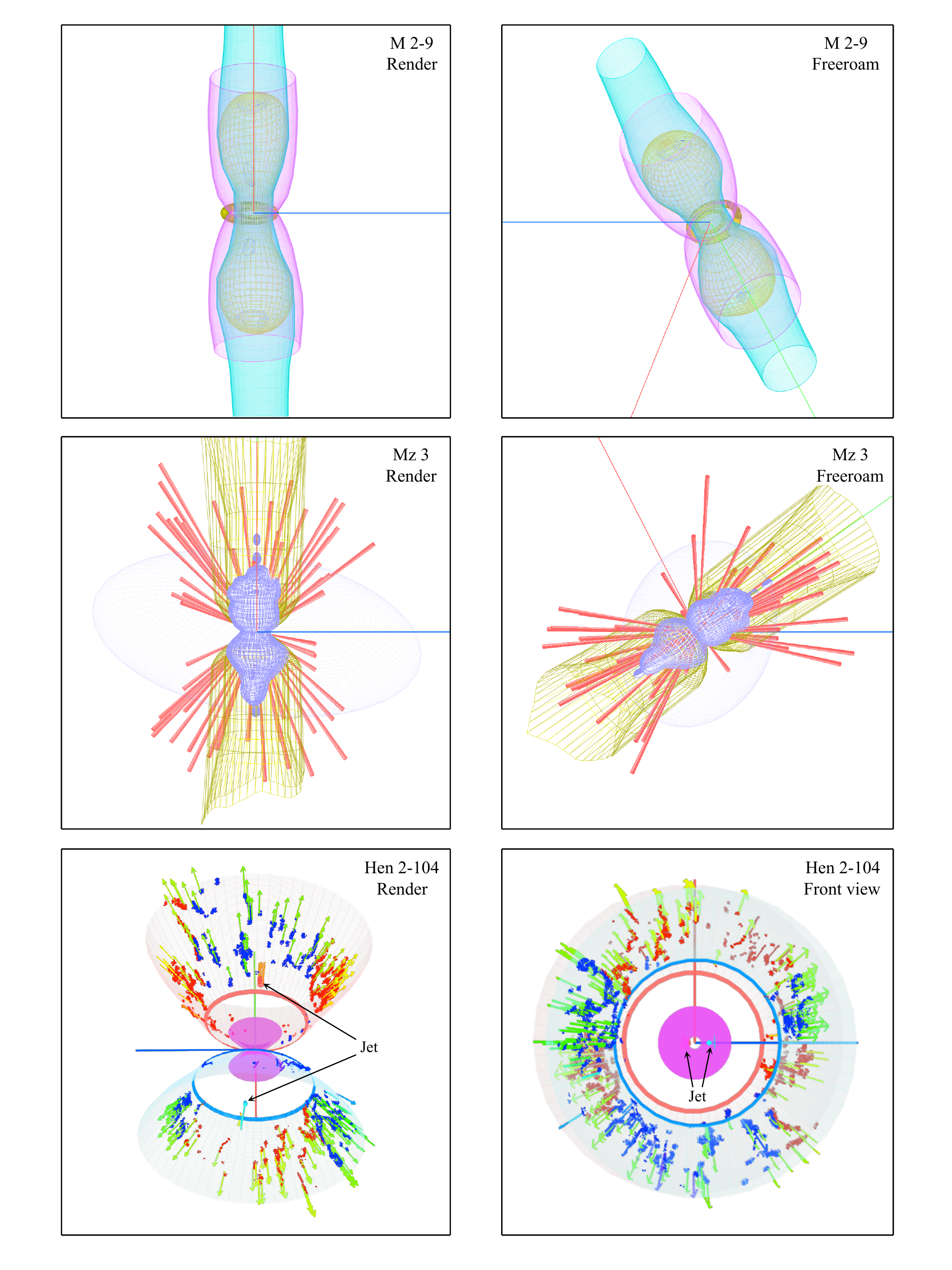}
\caption{Mesh images of M~2--9, Mz~3, and Hen 2--104 showing their components and sub-components. Shown on the left column are the render views for the images seen in Sects.~\ref{m2-9}, \ref{mz3}, and \ref{hen2-104}.}
\label{fig:henvelocityfield}
\end{figure*}
%

\end{document}